\documentclass[reprint,amsmath,amssymb,aps,superscriptaddress,floatfix]{revtex4-1}

\usepackage{graphicx}
\usepackage{dcolumn}
\usepackage{bm}
\usepackage{hyperref}
\usepackage{siunitx}
\usepackage{color}
\usepackage{xspace}

\DeclareSIUnit\kT{$k_B T$}
\DeclareSIUnit\dyne{dyne}
\DeclareSIUnit\molar{M}
\newcommand{\ee}{\mathrm{e}}

\newcommand{\ve}[1]{\bm{\mathbf{#1}}}
\newcommand{\veh}[1]{\bm{\mathbf{\hat{#1}}}}

\newcommand{\kT}{{k_B T}}

\newcommand{\nonum}{\nonumber\\}

\begin{document}

\title{A geometric attractor mechanism for self-organization of entorhinal grid modules}

\date{\today}

\author{Louis Kang}
\email[Corresponding author: ]{louis.kang@berkeley.edu}
\affiliation{David Rittenhouse Laboratories, University of Pennsylvania, Philadelphia, Pennsylvania, USA}
\affiliation{Redwood Center for Theoretical Neuroscience, University of California, Berkeley, Berkeley, California, USA}

\author{Vijay Balasubramanian}
\affiliation{David Rittenhouse Laboratories, University of Pennsylvania, Philadelphia, Pennsylvania, USA}

\begin{abstract}
  Grid cells in the medial entorhinal cortex (MEC) respond when an animal occupies a periodic lattice of ``grid fields'' in the environment. The grids are organized in modules with spatial periods, or scales, clustered around discrete values separated by ratios in the range 1.2--2.0. We propose a mechanism that produces this modular structure through dynamical self-organization in the MEC. In attractor network models of grid formation, the grid scale of a single module is set by the distance of recurrent inhibition between neurons. We show that the MEC forms a hierarchy of discrete modules if a smooth increase in inhibition distance along its dorso-ventral axis is accompanied by excitatory interactions along this axis. Moreover, constant scale ratios between successive modules arise through geometric relationships between triangular grids and have values that fall within the observed range. We discuss how interactions required by our model might be tested experimentally.
\end{abstract}

\maketitle

A grid cell has a spatially modulated firing rate that peaks when an animal reaches certain locations in its environment~\cite{Hafting:2005dp}. These locations of high activity form a regular triangular grid with a particular length scale and orientation in space. Every animal has many grid cells that collectively span a wide range of scales, with smaller scales enriched dorsally and larger scales ventrally along the longitudinal axis of the MEC~\cite{Stensola:2012gn}. Instead of being smoothly distributed, grid scales cluster around particular values and thus grid cells are partitioned into modules~\cite{Stensola:2012gn}. Consecutive pairs of modules have scale ratios in the range 1.2--2.0~\cite{Stensola:2012gn,Barry:2007gv,Krupic:2015gj}. Across animals, the average scale ratio is constant from one pair of modules to the next~\cite{Stensola:2012gn}. These observations underlie the possibility that the grid system favors a universal scale ratio in the range $1.4$~\cite{Stensola:2012gn} to $1.6$~\cite{Barry:2007gv,Krupic:2015gj}.

Encoding spatial information through grid cells with constant scale ratios is thought to provide animals with an efficient way of representing their position within an environment~\cite{Moser:2008hh,Fiete:2008dz, mathis2012optimal, Wei:2015hl,Stemmler:2015gc,Sanzeni:2016fg,Mosheiff:2017fj}. Moreover, periodic representations of space permit a novel mechanism for precise error correction against neural noise~\cite{Sreenivasan:2011fy} and are learned by machines seeking to navigate open environments~\cite{cueva2018emergence, banino2018vector}. These findings provide motivation for forming a modular grid system with a constant scale ratio, but a mechanism for doing so is unknown. Continuous attractor networks~\cite{Fuhs:2006fb,Burak:2009fx}, a leading model for producing grid cells, would currently require discrete changes in scales to be directly imposed as sharp changes in parameters, as would the oscillatory interference model~\cite{Burgess:2007fi,Hasselmo:2007cv} or hybrid models~\cite{Bush:2014iq}. In contrast, many sensory and behavioral systems have smooth tuning distributions, such as preferred orientation in visual cortex~\cite{Issa:2008de} and preferred head direction in the MEC~\cite{Taube:1990vf}. A self-organizing map model with stripe cell inputs~\cite{Grossberg:2012ih} and a firing rate adaptation model with place cell inputs~\cite{Urdapilleta:2017kn} can generate discrete grid scales, but their ratios are not constant or constant-on-average unless explicitly tuned.

Here, we present a simple extension of the continuous attractor model that adds excitatory connections between a series of attractor networks along the dorso-ventral axis of the MEC, accompanied by an increase in the distance of inhibition. The inhibition gradient drives an increase in grid scale along the MEC axis.  Meanwhile, the excitatory coupling discourages changes in grid scale and orientation unless they occur through geometric relationships with defined scale ratios and orientation differences. Competition between the effects of longitudinal excitation and lateral inhibition self-organizes the complete network into a discrete hierarchy of modules.   Certain grid relationships are geometrically stable, which makes them, and their associated scale ratios, insensitive to perturbations.   The precise ratios that appear depend on the balance between excitation and inhibition and how it varies along the MEC axis.  We show that sampling across a range of these parameters leads to a distribution of scale ratios that matches experiment and is, on average, constant from the smallest to the largest pair of modules.

Continuous attractors are a powerful general method for self-organizing neural dynamics. To our knowledge, our results are the first demonstration of a mechanism for producing a discrete hierarchy of modules in a continuous attractor system.

\section*{Results}

\subsection*{Standard grid cell attractors are not modular}

\begin{figure*}
	\includegraphics[width=\linewidth]{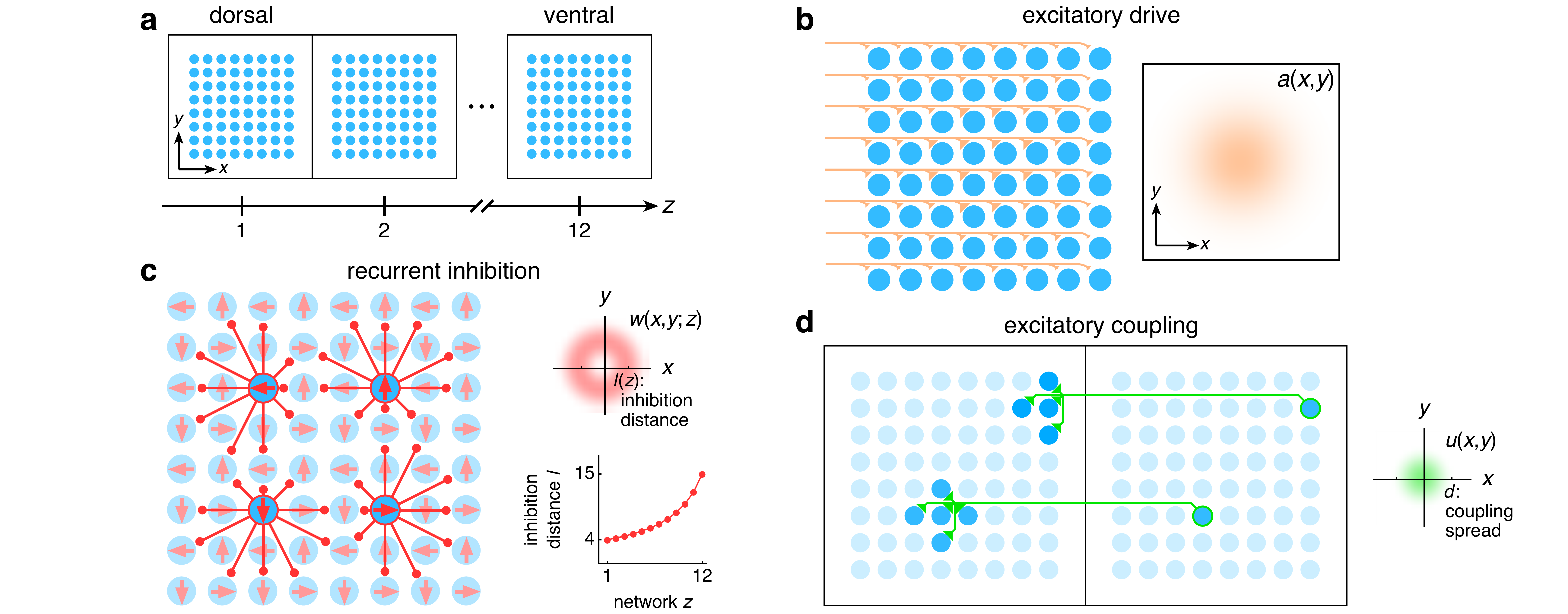}
	\caption{\label{fig:model}The entorhinal grid system as coupled 2D continuous attractor networks (\textbf{Methods}). (\textbf{a}) Each network $z$ corresponds to a region along the dorso-ventral MEC axis and contains a 2D sheet of neurons with positions $(x,y)$. (\textbf{b}) Neurons receive excitatory drive $a(x,y)$ that is greatest at the network center and decays toward the edges. (\textbf{c}) Neurons inhibit neighbors within the same network with a weight $w(x, y; z)$ that peaks at a distance of $l(z)$ neurons, which increases as a function of $z$. Each neuron has its inhibitory outputs shifted slightly in one of four preferred network directions and receives slightly more drive when the animal moves along its preferred spatial direction. (\textbf{d}) Each neuron at position $(x,y)$ in network $z$ excites neurons located within a spread $d$ of $(x,y)$ in network $z-1$.}
\end{figure*}

\begin{figure*}
	\includegraphics[width=\linewidth]{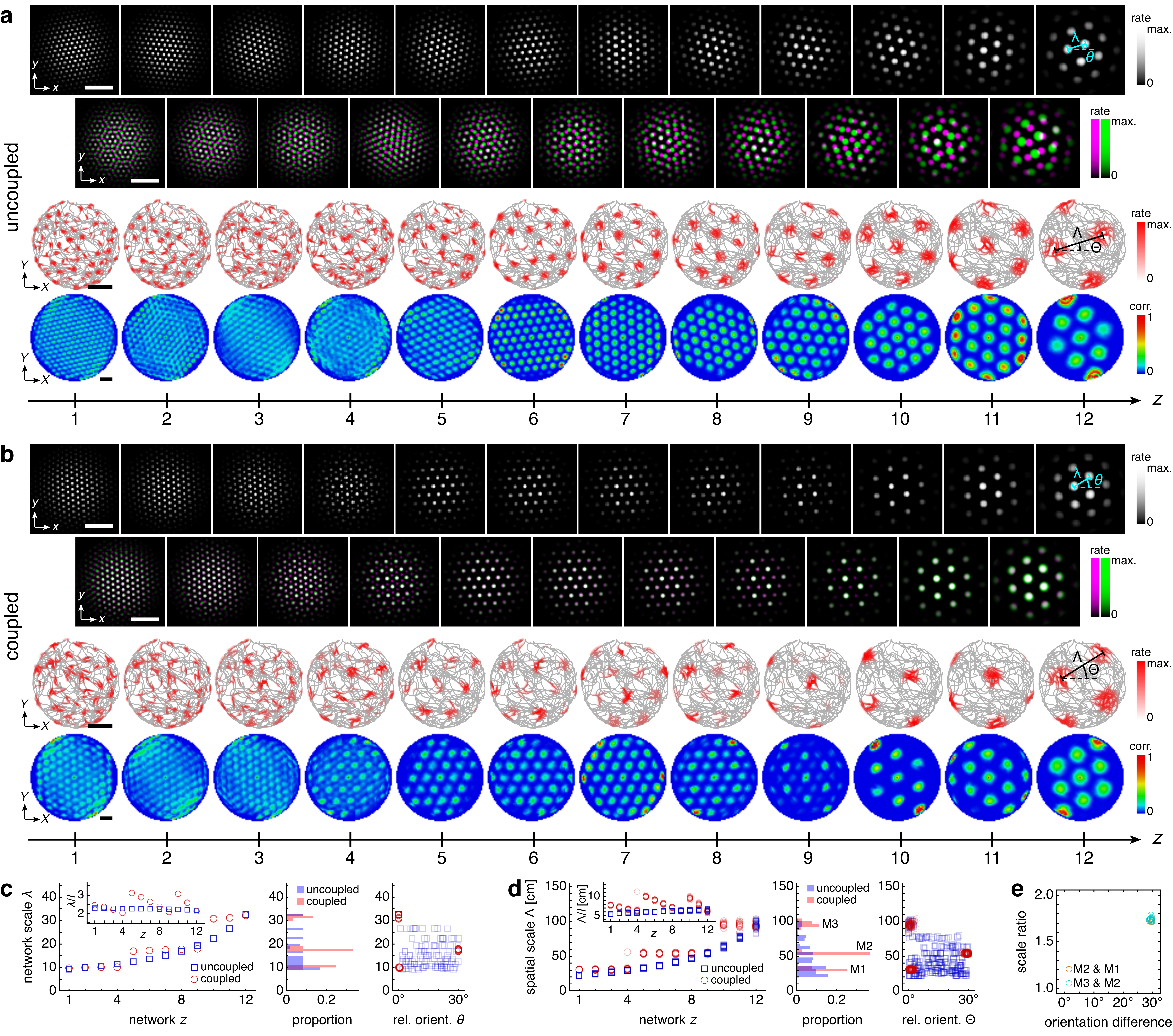}
  \caption{\label{fig:modules}Coupling can induce modularity with fixed scale ratios and orientation differences. (\textbf{a}) A representative simulation without coupling. Top row: network activities at the end of the simulation. Second row: activity overlays between adjacent networks depicted in the top row.  In each panel, the network with smaller (larger) $z$ is depicted in magenta (green), so white indicates activity in both networks. Third row: spatial rate map of a single neuron for each $z$ superimposed on the animal's trajectory. Bottom row: spatial autocorrelations of the rate maps depicted in the third row. White scale bars, 50 neurons. Black scale bars, \SI{50}{\cm}. (\textbf{b}) Same as \textbf{a} but for a representative simulation with coupling. (\textbf{c}--\textbf{e}) Data from 10 replicate uncoupled and coupled simulations. (\textbf{c}) Left: network grid scales $\lambda(z)$. For each network, there are 10 closely spaced red circles and 10 closely spaced blue squares corresponding to replicate simulations. Inset: $\lambda(z)$ divided by the inhibition distance $l(z)$. Middle: histogram for $\lambda$ collected across all networks. Right: network grid orientations $\theta$ relative to the network in the same simulation with largest scale. (\textbf{d}) Left: spatial grid scales $\Lambda(z)$. For each $z$, there are up to 30 red circles and 30 blue squares corresponding to 3 neurons recorded during each simulation. Inset: $\Lambda(z)$ divided by the inhibition distance $l(z)$. Middle: histogram for $\Lambda$ collected across all networks. In the coupled model, grid cells are clustered into three modules. Right: spatial grid orientations $\Theta$ relative to the grid cell in the same simulation with largest scale. (\textbf{e}) Spatial scale ratios and orientation differences between adjacent modules for the coupled model. Standard parameter values provided in \textbf{Table~\ref{tab:params}}.}
\end{figure*}

We assemble a series of networks along the longitudinal MEC axis, numbering them $z = 1, 2, \ldots, 12$ from dorsal to ventral (\textbf{Fig.~\ref{fig:model}a}). Each network contains the standard 2D continuous attractor architecture of the Burak-Fiete model~\cite{Burak:2009fx}. Namely, neurons are arranged in a 2D sheet with positions $(x,y)$, receive broad excitatory drive (Ref.~\onlinecite{Bonnevie:2013eu} and \textbf{Fig.~\ref{fig:model}b}), and inhibit one another at a characteristic separation on the neural sheet (\textbf{Fig.~\ref{fig:model}c}; see \textbf{Methods} for a complete description). This inhibition distance $l$ is constant within each network but increases from one network to the next along the longitudinal axis of the MEC. With these features alone, the population activity in each network self-organizes into a triangular grid whose lattice points correspond to peaks in neural activity (first row of \textbf{Fig.~\ref{fig:modules}a}). Importantly, the scale of each network's grid, which we call $\lambda(z)$, is proportional to that network's inhibition distance $l(z)$ (``uncoupled'' simulations in \textbf{Fig.~\ref{fig:modules}c}). Also, network grid orientations $\theta$ show no consistent pattern across scales and among replicate simulations with different random initial firing rates.

Following the standard attractor model~\cite{Burak:2009fx}, the inhibitory connections in each network are slightly modulated by the animal's velocity such that the population activity pattern of each network translates proportionally to animal motion at all times (\textbf{Methods}). This modulation allows each network to encode the animal's displacement through a process known as path-integration, and projects the network grid pattern onto spatial rate maps of single neurons. That is, a recording of a single neuron over the course of an animal trajectory would show high activity in spatial locations that form a triangular grid with scale $\Lambda$ (third row of \textbf{Fig.~\ref{fig:modules}a}). Moreover, $\Lambda(z)$ for a neuron from network $z$ is proportional to that network's population grid scale $\lambda(z)$, and thus also proportional to its inhibition distance $l(z)$ (uncoupled simulations in \textbf{Fig.~\ref{fig:modules}d}). To be clear, we call $\Lambda$ the ``spatial scale''; it corresponds to a single neuron's activity over the course of a simulation and has units of physical distance in space. By contrast, $\lambda$, the ``network scale'' described above, corresponds to the population activity at a single time and has units of separation on the neural sheet. Similarly, $\Theta(z)$ describes the orientation of the spatial grid of a single neuron in the network $z$; we call $\Theta$ the ``spatial orientation''. Like the network orientations $\theta$ discussed above, spatial orientations of grids show no clustering (uncoupled simulations in \textbf{Fig.~\ref{fig:modules}d}).

With an inhibition distance $l(z)$ that increases gradually from one network to the next (\textbf{Fig.~\ref{fig:model}c}), proportional changes in network and spatial scales $\lambda(z)$ and $\Lambda(z)$ lead to a smooth distribution of grid scales (uncoupled simulations in \textbf{Fig.~\ref{fig:modules}c},~\textbf{d}). To reproduce the experimentally observed jumps in grid scale between modules, the inhibition length would also have to undergo discrete, sharp jumps between certain adjacent networks. A further mechanism would be needed to enforce the preferred orientation differences that are observed between modules. In summary, a grid system created by disjoint attractor networks will not self-organize into modules.

\subsection*{Coupled attractor networks produce modules}

Module self-organization can be achieved with one addition to the established features listed above: we introduce excitatory connections from each neuron to those in the preceding network with approximately corresponding neural sheet positions (\textbf{Fig.~\ref{fig:model}d}; see \textbf{Methods} for a complete description). That is, a neuron in network $z$ (more ventral) with position $(x,y)$ will excite neurons in network $z-1$ (more dorsal) with positions that are within a distance $d$ of position $(x,y)$. In other words, the distance $d$ is the ``spread'' of excitatory connections, and we choose a constant value across all networks comparable to the inhibition distance $l(z)$. Similar results are obtained with dorsal-to-ventral or bidirectional excitatory coupling (below) or with a spread $d(z)$ that increases with the inhibition distance $l(z)$ (\textbf{Supp.\@ Fig.~1}).

The self-organization of triangular grids in the neural sheet and the faithful path-integration that projects these grids onto single neuron spatial rate maps persist after introduction of inter-network coupling (\textbf{Fig.~\ref{fig:modules}b}). Network and spatial scales $\lambda(z)$ and $\Lambda(z)$ still increase from network $z = 1$ (dorsal) to network $z = 12$ (ventral). Yet, \textbf{Fig.~\ref{fig:modules}c},~\textbf{d} shows that for the coupled model, these scales exhibit plateaus that are interrupted by large jumps, disrupting their proportionality to inhibition distance $l(z)$, which is kept identical to that of the uncoupled system (\textbf{Fig.~\ref{fig:model}c}). Collecting scales across all networks illustrates that they cluster around certain values in the coupled system while they are smoothly distributed in the uncoupled system.  We identify these clusters with modules M1, M2, and M3 of increasing scale.  Note that multiple networks at various depths $z$ can belong to the same module.   Moreover, coupling causes grid cells that cluster around a certain scale to also cluster around a certain orientation (\textbf{Fig.~\ref{fig:modules}c},~\textbf{d}), as seen in experiment~\cite{Stensola:2012gn}. The uncoupled system does not demonstrate co-modularity of orientation with scale, i.e., two networks with similar grid scales need not have similar orientations unless this is imposed by an external constraint.

In summary, excitatory coupling between grid attractor networks dynamically induces discreteness in grid scales that is co-modular with grid orientation, as observed experimentally~\cite{Stensola:2012gn}, and as needed for even coverage of space by the grid map~\cite{Sanzeni:2016fg}.

\subsection*{Modular geometry is determined by lattice geometry}

\begin{figure}
	\includegraphics[width=\linewidth]{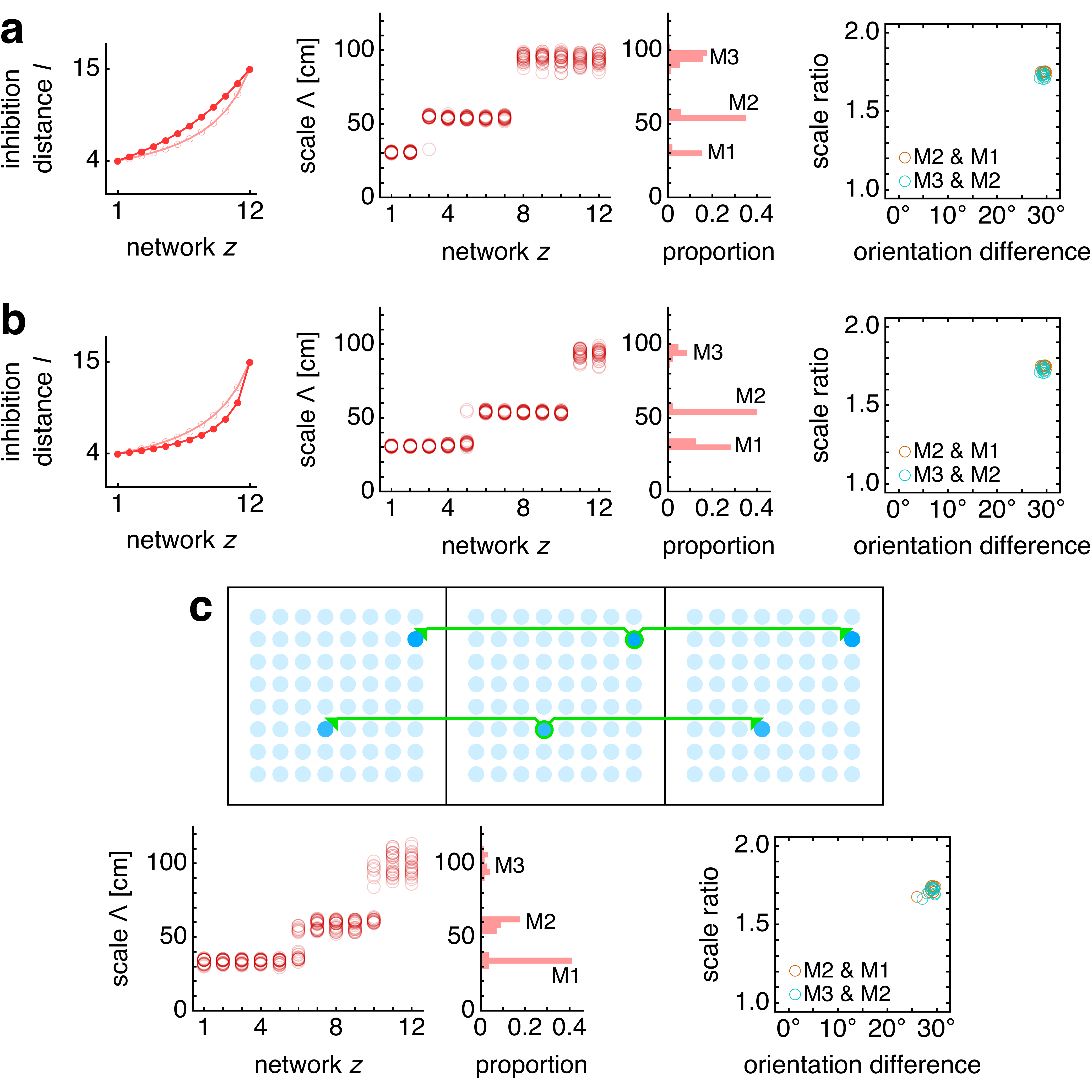}
	\caption{\label{fig:same}Modules produced by commensurate lattices are robust to parameter perturbations. Data from 10 replicate simulations in each subfigure. (\textbf{a}) Left: we use a less concave inhibition distance profile $l(z)$ (dark filled circles) compared to \textbf{Fig.~\ref{fig:model}c} (light empty circles). Middle: spatial grid scales exhibit modules when collected in a histogram across networks. Right: modules have the same scale ratios and orientation differences as in \textbf{Fig.~\ref{fig:modules}e}. (\textbf{b}) Same as \textbf{a}, but with a more concave $l(z)$. (\textbf{c}) Simulations with bidirectional point-to-point coupling instead of the unidirectional spread coupling in \textbf{Fig.~\ref{fig:model}d}. Top: schematic of the neuron at position $(x,y)$ in network $z$ exciting only the neuron at $(x,y)$ in networks $z-1$ and $z+1$. Bottom left/right: same as middle/right in \textbf{a}. In \textbf{a}, inhibition distance exponent $l_\textrm{exp} = 0$. In \textbf{b}, $l_\textrm{exp} = -2$. In \textbf{c}, coupling spread $d = 1$ and coupling strength $u_\textrm{mag} = 0.4$ in both directions. Other parameter values are in \textbf{Table~\ref{tab:params}}.}
\end{figure}

Not only does excitatory coupling produce modules, it can do so with consistent scale ratios and orientation differences. For the coupled system depicted in \textbf{Fig.~\ref{fig:modules}}, scale ratios and orientation differences between pairs of adjacent modules consistently take values $\num{1.74(2)}$ and $\SI{29.5(4)}{\degree}$, respectively (mean $\pm$ s.d.; \textbf{Fig.~\ref{fig:modules}e}). If we perturb the inhibition distance profile $l(z)$ by making it less or more concave, these scale ratios and orientation differences are unchanged (\textbf{Fig.~\ref{fig:same}a},~\textbf{b}). Concavity only affects the number of grid cells in each module, which can be tuned to match experimental observations. The same scale ratios and orientation differences also persist after changes to the directionality and spread of excitatory connections. For example, we replace the ventral-to-dorsal connections with bidirectional coupling and decrease the coupling spread $d$ such that a neuron in network $z$ excites only a single neuron in both networks $z-1$ and $z+1$; scale ratios and orientation differences remain at $1.7$ and $\SI{30}{\degree}$, respectively (\textbf{Fig.~\ref{fig:same}c}). Representative network activities and single neuron rate maps for these simulations are provided in \textbf{Supp.\@ Fig.~2}. Data for simulations with only dorsal-to-ventral connections are provided in \textbf{Supp.\@ Fig.~3}; they also exhibit the same scale ratios and orientation differences.

We can intuitively understand this precise modularity through the competition between lateral inhibition within networks and longitudinal excitation across networks. In the uncoupled system, grid scales decrease proportionally as the inhibition distance $l(z)$ decreases from $z = 12$ to $z = 1$. However, coupling causes areas of high activity in network $z$ to preferentially excite corresponding areas in network $z-1$, which encourages adjacent networks to share the same grid pattern. Thus, coupling adds rigidity to the system and provides an opposing ``force'' against the changing inhibition distance that attempts to drive changes in grid scale. This rigidity produces the plateaus in network and spatial scales $\lambda(z)$ and $\Lambda(z)$ that delineate modules across multiple networks.

At interfaces between modules, coupling can no longer fully oppose the changing inhibition distance, and the grid pattern changes. However, the rigidity fixes a geometric relationship between the grid patterns of the two networks spanning the interface. In the coupled system of \textbf{Fig.~\ref{fig:modules}}, module interfaces occur between networks $z = 4$ and $5$ and between $z = 9$ and $10$. The network population activity overlays of \textbf{Fig.~\ref{fig:modules}b} reveal overlap of many activity peaks at these interfaces. However, the more dorsal network (with smaller $z$) at each interface contains additional small peaks between the shared peaks.  In this way, adjacent networks still share many corresponding areas of high activity, as favored by coupling, but the grid scale changes, as favored by a changing inhibition distance. Pairs of grids whose lattice points demonstrate regular registry are called \emph{commensurate} lattices~\cite{chaikinlubensky} and have precise scale ratios and orientation differences, here respectively $\sqrt{3} \approx 1.7$ and $\SI{30}{\degree}$, which match the results in \textbf{Figs.~\ref{fig:modules}e} and \textbf{\ref{fig:same}}.

In summary, excitatory coupling can compete against a changing inhibition distance to produce a rigid grid system whose ``fractures'' exhibit stereotyped commensurate lattice relationships. These robust geometric relationships lead to discrete modules with fixed scale ratios and orientation differences.

In our model, commensurate lattice relationships naturally lead to field-to-field firing rate variability in single neuron spatial rate maps (for example, $z = 8$ in the third row of \textbf{Fig.~\ref{fig:modules}b}), another experimentally observed feature of the grid system~\cite{Ismakov:2017jj,Dunn:2017jk}. At interfaces between two commensurate lattices, only a subset of population activity peaks in the grid of smaller scale overlap with, and thus receive excitation from, those in the grid of larger scale. The network with smaller grid scale will contain activity peaks of different magnitudes; this heterogeneity is then projected onto the spatial rate maps of its neurons.  

\subsection*{Excitation-inhibition balance sets lattice geometry}

\begin{figure*}
	\includegraphics[width=\linewidth]{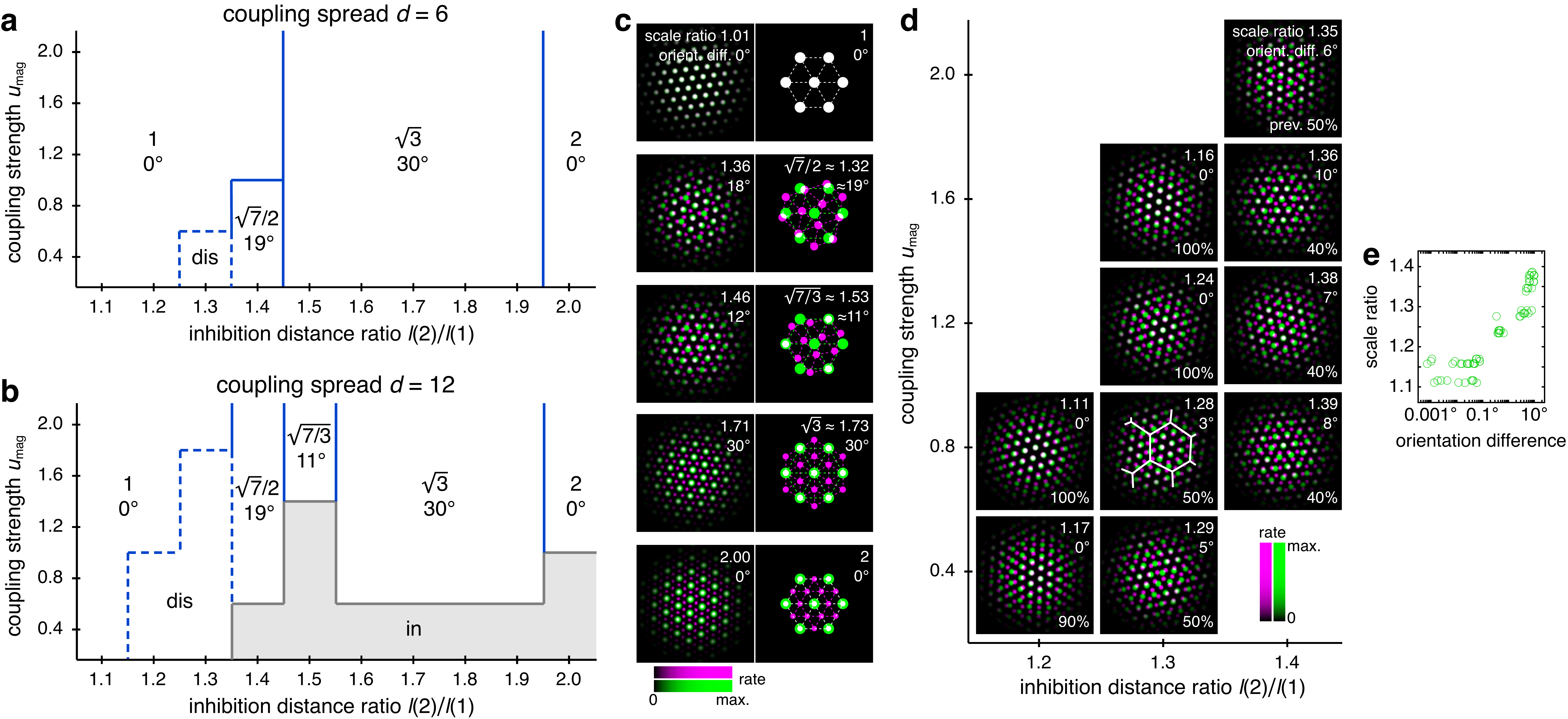}
  \caption{\label{fig:phase}Diverse lattice relationships emerge over wide ranges in simulation parameters. In models with only two networks $z = 1$ and $2$, we vary the coupling strength $u_\textrm{mag}$ and the ratio of inhibition distances $l(2)/l(1)$ for two different coupling spreads $d$. (\textbf{a},~\textbf{b}) Approximate phase diagrams based on 10 replicate simulations for each set of parameters, with the mean of $l(1)$ and $l(2)$ fixed to be 9. The most frequently occurring scale ratio and orientation difference are indicated for each region; coexistence between multiple lattice relationships may exist at drawn boundaries. (\textbf{a}) Phase diagram for small coupling spread $d = 6$. Solid lines separate four regions with different commensurate lattice relationships labeled by scale ratio and orientation difference, and dotted lines mark one region of discommensurate lattice relationships. (\textbf{b}) Phase diagram for large coupling spread $d = 12$. There are five different commensurate regions, a discommensurate region, as well as a region containing incommensurate lattices (gray). (\textbf{c}) Network activity overlays for representative observed (left) and idealized (right) commensurate relationships. Numbers at the top right of each image indicate network scale ratios $\lambda(2)/\lambda(1)$ and orientation differences $\theta(2) - \theta(1)$.  Networks $z = 1$ and $2$ in magenta and green, respectively, so white indicates activity in both networks. (\textbf{d}) Expanded region of \textbf{b} displaying discommensurate lattice statistics. For each set of parameters, a representative overlay for the most prevalent discommensurate lattice relationship is shown. The number in the lower right indicates the proportion of replicate simulations with scale ratio within 0.01 and orientation difference within \SI{3}{\degree} of the values shown at top right. In one overlay, discommensurations are outlined by white lines. (\textbf{e}) The discommensurate relationships described in \textbf{d} demonstrate positive correlation between scale ratio and the logarithm of orientation difference (Pearson's $\rho = 0.91$). Parameter values provided in \textbf{Supp.\@ Info.}}
\end{figure*}

Adjusting the balance between excitatory coupling and a changing inhibition distance produces other commensurate lattice relationships, each of which enforces a certain scale ratio and orientation difference. To explore this competition systematically, we use a smaller coupled model with just two networks, $z = 1$ and $2$, and vary three parameters: the coupling spread $d$, the coupling strength $u_\textrm{mag}$, and the ratio of inhibition distances between the two networks $l(2)/l(1)$ (\textbf{Supp.\@ Info.}). For each set of parameters, we measure network scale ratios and orientation differences produced by multiple replicate simulations (\textbf{Supp.\@ Fig.~4}). We find that as the excitation-inhibition balance is varied by changing $u_\textrm{mag}$ and $l(2)/l(1)$, a number of discretely different relationships appear, which can be summarized in ``phase diagrams'' (\textbf{Fig.~\ref{fig:phase}a},~\textbf{b}).

In many regions of the phase diagrams, these lattice relationships are commensurate, each with a characteristic scale ratio and orientation difference (\textbf{Fig.~\ref{fig:phase}c}). When parameters are chosen near a boundary between two regions, replicate simulations may adopt either lattice relationship or occasionally be trapped in other metastable relationships due to variations in random initial conditions (\textbf{Supp.\@ Fig.~4}). At larger $u_\textrm{mag}$ in both phase diagrams, there are fewer regions as $l(2)/l(1)$ varies because a higher excitatory coupling strength provides more rigidity against gradients in inhibition distance (\textbf{Fig.~\ref{fig:phase}a},~\textbf{b}). However, a larger coupling spread $d$ would cause network $z = 2$ to excite a broader set of neurons in network $z = 1$, softening the rigidity imposed by coupling and producing a wider variety of lattices in \textbf{Fig.~\ref{fig:phase}b} than \textbf{Fig.~\ref{fig:phase}a}. Also in \textbf{Fig.~\ref{fig:phase}b}, when excitation is weak and approaching the uncoupled limit, there is a noticeable region dominated by \emph{incommensurate} lattices, in which the two grids lack consistent registry or relative orientation, and grid scale is largely determined by inhibition distance (\textbf{Supp.\@ Fig.~4}).

\textbf{Figure~\ref{fig:phase}b} also contains a larger region of \emph{discommensurate} lattices (although strictly speaking, in condensed matter physics, they would be termed commensurate lattices with discommensurations~\cite{chaikinlubensky}). Discommensurate networks have closely overlapping activities in certain areas that are separated by a mesh of regions lacking overlap called discommensurations (\textbf{Fig.~\ref{fig:phase}d}). They exhibit ranges of scale ratios 1.1--1.4 and orientation differences \SI{0}{\degree}--\SI{10}{\degree} that ultimately arise from a single source: the density of discommensurations, whose properties can also be explained through excitation-inhibition competition. Stronger coupling drives more activity overlap, which favors sparser discommensurations and lowers the scale ratio and orientation difference. However, a larger inhibition distance ratio drives the two networks to differ more in grid scale, which favors denser discommensurations. To better accommodate the discommensurations, grids rotate slightly as observed previously in a crystal system~\cite{Wilson:1990cj}. \textbf{Figure~\ref{fig:phase}e} confirms that scale ratios and orientation differences vary together as the discommensuration density changes. 

Thus, by changing the balance between excitation and inhibition, a two-network model yields geometric lattice relationships with various scale ratios and corresponding orientation differences. All of the commensurate relationships (\textbf{Fig.~\ref{fig:phase}c}) and almost the entire range of discommensurate relationships (\textbf{Fig.~\ref{fig:phase}d}) have scale ratios that fall in the range of experimental measurements, which is roughly 1.2--2.0~\cite{Stensola:2012gn,Barry:2007gv,Krupic:2015gj}.

\subsection*{Discommensurate lattices produce distinct modular geometries but with more variation}

\begin{figure*}
	\includegraphics[width=\linewidth]{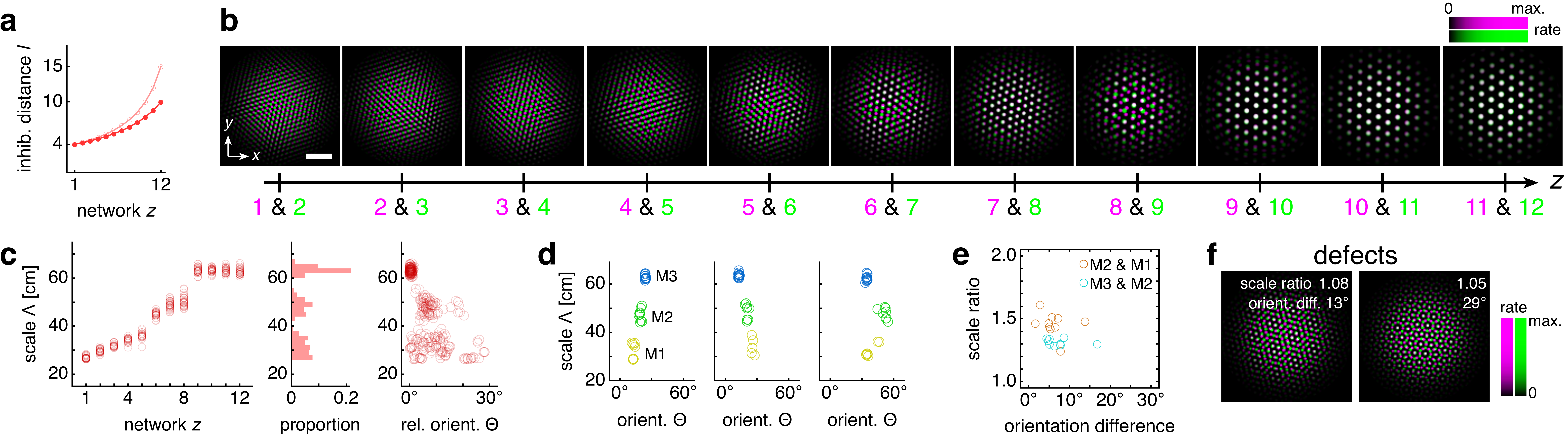}
  \caption{\label{fig:dis}Discommensurate lattice relationships can produce realistic modules. (\textbf{a}) We use a shallower inhibition distance profile $l(z)$ (dark filled circles) compared to \textbf{Fig.~\ref{fig:model}c} (light empty circles). (\textbf{b}) Representative activity overlays between adjacent networks $z$ in magenta and green, so white indicates activity in both networks. Scale bar, 50 neurons. (\textbf{c--e}) Data from 10 replicate simulations. (\textbf{c}) Left: spatial grid scales $\Lambda(z)$. For each network, there are up to 30 red circles corresponding to 3 neurons recorded during each simulation. Middle: histogram for $\Lambda$ collected across all networks. Right: spatial orientations $\Theta$ relative to the grid cell in the same simulation with largest scale. (\textbf{d}) Clustering of spatial scales and orientations for 3 representative simulations. Due to 6-fold lattice symmetry, orientation is a periodic variable modulo $\SI{60}{\degree}$. Different colors indicate separate modules. (\textbf{e}) Spatial scale ratios and orientation differences between adjacent modules. (\textbf{f}) Representative activity overlays demonstrating defects with low activity overlap. Maximum inhibition distance $l_\textrm{max} = 10$, coupling spread $d = 12$. We use larger network size $n \times n = 230 \times 230$ to allow for discommensurate relationships whose periodicities span longer distances on the neural sheets. Other parameter values are in \textbf{Table~\ref{tab:params}}.}
\end{figure*}

As mentioned above, discommensurate lattices have a range of allowed geometries (\textbf{Fig.~\ref{fig:phase}d},~\textbf{e}), but they still produce modules in a full 12-network grid system with a preferred scale ratio and orientation difference. However, these values do not cluster as strongly as they do for a commensurate relationship, which is geometrically precise.

The phase diagrams of \textbf{Fig.~\ref{fig:phase}} provide guidance for modifying a 12-network system that exhibits a $[\sqrt{3}, \SI{30}{\degree}]$ relationship to produce discommensurate relationships instead. We make the inhibition distance profile $l(z)$ shallower (\textbf{Fig.~\ref{fig:dis}a}) and increase the coupling spread $d$ by 50\%. Network activity overlays of these new simulations reveal grids obeying discommensurate relationships (\textbf{Fig.~\ref{fig:dis}b}), which are projected onto single neuron spatial rate maps through faithful path-integration (\textbf{Supp.\@ Fig.~5}). Across replicate simulations with identical parameter values but different random initial firing rates, the discommensurate system demonstrates greater variation in scale and orientation (\textbf{Fig.~\ref{fig:dis}c}) than the commensurate systems of \textbf{Figs.~\ref{fig:modules}} and \textbf{\ref{fig:same}}. Nevertheless, analysis of each replicate simulation reveals clustering with well-defined modules (\textbf{Fig.~\ref{fig:dis}d} and \textbf{Supp.\@ Fig.~5}). These modules have scale ratio $\num{1.39(10)}$ and orientation difference $\SI{6.7(35)}{\degree}$ (mean $\pm$ s.d.; \textbf{Fig.~\ref{fig:dis}e}). The preferred scale ratio agrees well with the mean value observed experimentally in~\cite{Stensola:2012gn}.

Conceptually, we can interpret the greater spread of scales and orientations in terms of coupling rigidity. Excitatory coupling, especially when the spread is larger, provides enough rigidity in the discommensurate system to cluster scale ratios and orientation differences but not enough to prevent variations in these values. The degree of variability observed in \textbf{Fig.~\ref{fig:dis}c},~\textbf{d} appears consistent with experimental measurements, which also demonstrate spread~\cite{Stensola:2012gn,Barry:2007gv}.

A few module pairs in \textbf{Fig.~\ref{fig:dis}e} exhibit a large orientation difference ${>}\SI{10}{\degree}$. This is not expected from a discommensurate relationship, and indeed, inspecting the network activities reveals adjacent networks trapped in a relationship with low activity overlap and large orientation difference (\textbf{Fig.~\ref{fig:dis}f}). In the context of a grid system that otherwise obeys commensurate or discommensurate geometries containing more overlap, we call this less common relationship a ``defect.'' We distinguish between these relationships and the incommensurate lattices discussed above, which also have low activity overlap. Defects arise when the excitatory coupling is strong,  and incommensurate lattices arise when this coupling is weak. Also, defects have smaller scale ratios ${<}1.1$ and larger orientation differences ${>}\SI{10}{\degree}$, whereas incommensurate lattices have larger scale ratios ${>}1.3$ and any orientation difference (\textbf{Fig.~\ref{fig:phase}b} and \textbf{Supp.\@ Fig.~4}).

Thus, networks governed by discommensurate relationships also cluster into modules with a preferred scale ratio and orientation difference within the experimental range~\cite{Stensola:2012gn,Krupic:2015gj}. Due to lower coupling rigidity compared to commensurate grid systems, they exhibit increased variability and occasional defects across replicate simulations.

\subsection*{A diversity of lattice geometries maintains constant-on-average scale ratios}

\begin{figure}
	\includegraphics[width=\linewidth]{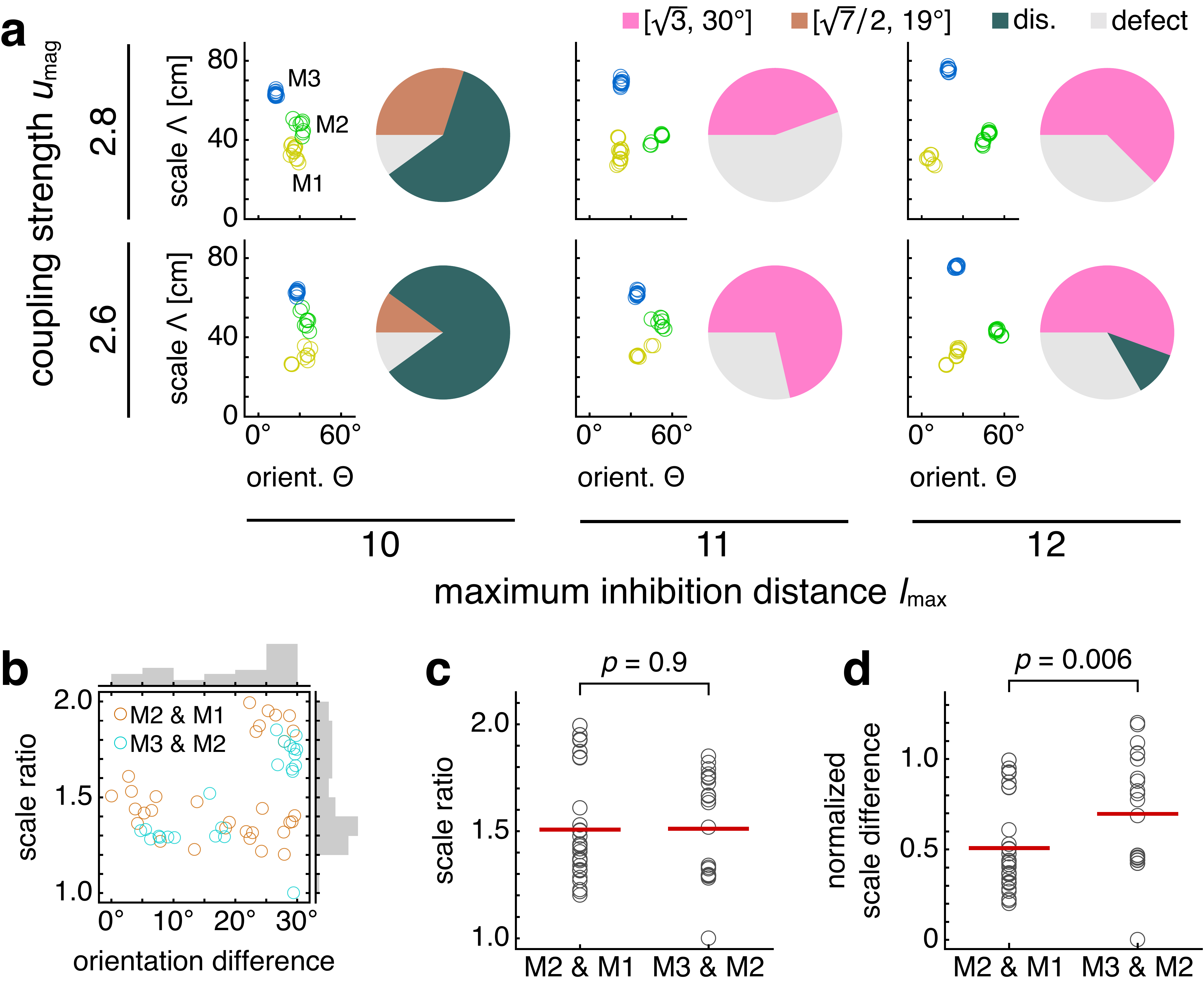}
  \caption{\label{fig:range}Simulations spanning different parameters contain diversity in lattice relationships, but average scale ratios are still constant between module pairs. Data from 5 replicate simulations for each set of parameters. (\textbf{a}) Clustering of spatial scales and orientations for one representative simulation (left) and lattice relationship distribution across all pairs of adjacent modules (right) for each set of parameters. (\textbf{b}) Spatial scale ratios and orientation differences between adjacent modules with respective histograms to the right and above. Scale ratios and orientation differences exhibit positive rank correlation (Spearman's $\rho = 0.44$, $p = 0.001$). (\textbf{c}) Spatial scale ratios. Means indicated by lines. Medians compared through the Mann-Whitney $U$ test with reported $p$-value. (\textbf{d}) Spatial scale differences normalized by the scale of the first module (M1) in each simulation. Same interpretation of lines and $p$-value as in \textbf{c}. The $u_\textrm{mag} = 2.6$ and $l_\textrm{max} = 10$ data are taken from simulations in \textbf{Fig.~\ref{fig:dis}}. Some simulations produced only two modules M1 and M2; one simulation produced four modules, and M4 was excluded from further analysis (\textbf{Supp.\@ Fig.~6}). Coupling spread $d = 12$ and network size $n \times n = 230 \times 230$. Other parameter values are in \textbf{Table~\ref{tab:params}}.}
\end{figure}

So far, each set of 12-network simulations contained replicates with identical parameter values and exhibited a single dominant lattice relationship. We now present results with different parameter values to imitate biological network variability across animals. This procedure leads to modules with different commensurate and discommensurate relationships (\textbf{Fig.~\ref{fig:range}a} and \textbf{Supp.\@ Fig.~6}). There is no longer a single preferred scale ratio or orientation difference (\textbf{Fig.~\ref{fig:range}b}), but patterns emerge due to the predominance of discommensurate and commensurate relationships. Recall from \textbf{Fig.~\ref{fig:dis}e} that discommensurate module pairs exhibit scale ratios ${\approx}1.4$ and orientation differences ${\approx}\SI{7}{\degree}$. Combined with $[\sqrt{3} \approx 1.7, \SI{30}{\degree}]$ module pairs we find a bimodal distribution of orientation differences around $\SI{7}{\degree}$ and $\SI{30}{\degree}$, consistent with experimental data~\cite{Krupic:2015gj}, and positive correlation between scale ratio and orientation difference. Modules with low scale ratio but high orientation difference decrease this correlation; they arise from defects (\textbf{Fig.~\ref{fig:dis}f}).

Scale ratios across the network variations span a range of values, but their averages are constant across module pairs. That is, the median scale ratio does not change between the pair of modules with smaller scales and the larger pair (\textbf{Fig.~\ref{fig:range}c}). Similarly, mean values are respectively \num{1.52(5)} and \num{1.53(5)} (mean $\pm$ s.e.m.) for module pairs M2 \& M1 and M3 \& M2. Combining data from both module pairs gives scale ratio \num{1.52(3)} (mean $\pm$ s.e.m.), which agrees well with the mean value of 1.56 from Ref.~\onlinecite{Krupic:2015gj}. Reference~\onlinecite{Stensola:2012gn} reports a slightly smaller mean value of \num{1.42(17)} (mean $\pm$ s.d.; re-analyzed by Ref.~\onlinecite{Wei:2015hl}), but its broad distribution of scale ratios overlaps considerably with ours. Moreover, we find that the normalized scale \emph{difference} does change its median across module pairs (\textbf{Fig.~\ref{fig:range}d}). This result that scale ratios are constant on average but scale differences are not matches experiment~\cite{Stensola:2012gn}.

Thus, although our model can produce modules with fixed scale ratios, allowing for a range of network parameters also produces modules with a range of scale ratios. Nevertheless, the scale ratio averaged over these parameters is still constant across module pairs, a key feature of the grid system that holds even if scales are not governed by a universal ratio~\cite{Stensola:2012gn}.

\subsection*{Testing for coupling with a mock lesion experiment}

\begin{figure*}
	\includegraphics[width=\linewidth]{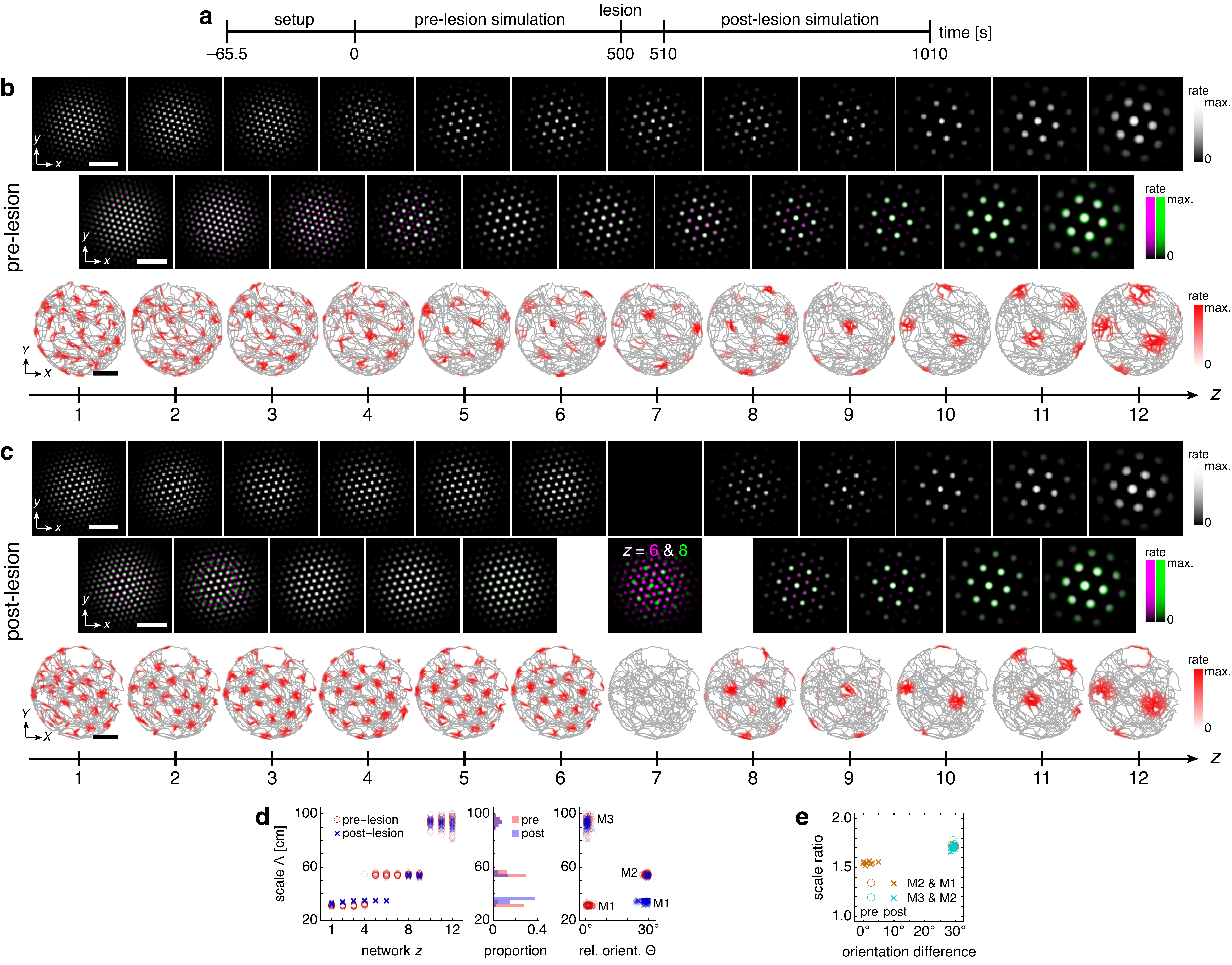}
  \caption{\label{fig:lesion}Lesioning a network changes grid scales and orientations of more dorsal networks. (\textbf{a}) Lesion protocol. (\textbf{b}) A representative simulation before the lesion. Top row: network activities at the end of the pre-lesion simulation. Second row: activity overlays between adjacent networks depicted in the top row. In each panel, the network with smaller (larger) $z$ is depicted in magenta (green), so white indicates activity in both networks. Third row: spatial rate map of a single neuron for each $z$ superimposed on the animal's trajectory. White scale bars, 50 neurons. Black scale bars, \SI{50}{\cm}. (\textbf{c}) Same as \textbf{b} but after the lesion. Spatial rate maps are recorded from the same neurons as in \textbf{b}. (\textbf{d},~\textbf{e}) Data from 10 replicate simulations. (\textbf{d}) Left: spatial grid scales $\Lambda(z)$ before and after the lesion. Middle: histogram for $\Lambda$ collected across all networks. Right: spatial orientations $\Theta$ relative to the grid cell in the same simulation with largest scale. (\textbf{e}) Spatial scale ratios and orientation differences between adjacent modules. Standard parameter values provided in \textbf{Table~\ref{tab:params}}.}
\end{figure*}

Excitatory coupling locks networks into scales and orientations imposed by more ventral networks. Disrupting the coupling frees networks from this rigidity, which can change scales and orientations far from the disruption. We demonstrate this effect by inactivating one network $z = 7$ midway through the simulation (\textbf{Fig.~\ref{fig:lesion}a}). This corresponds experimentally to disrupting all excitatory connections at one location along the dorsoventral MEC axis.

After the lesion, grid cells ventral to the lesion location ($z \geq 8$) are unaffected, but those dorsal to the lesion location ($z \leq 6$) change scale and orientation and form a single module (\textbf{Fig.~\ref{fig:lesion}b}--\textbf{d}). Network $z = 6$ is no longer constrained by larger grids of more ventral networks, so its scale decreases. The coupling that remains from $z = 6$ to $1$ then rigidly propagates the new grid down to network $z = 1$. This post-lesion module M1 has larger scale and \SI{30}{\degree} orientation difference compared to the pre-lesion M1; these changes also appear as corresponding changes in the scale ratio and orientation difference between modules M2 and M1 (\textbf{Fig.~\ref{fig:lesion}e}).

Immediate changes in grid scale and/or orientation observed at one location along the longitudinal MEC axis due to a lesion at another location would strongly support the presence of the excitatory coupling predicted by our model. Moreover, the anatomical distribution of the changes would indicate the directionality of coupling; those in grid cells dorsal to the lesion would indicate ventral-to-dorsal coupling and those ventral to the lesion would indicate dorsal-to-ventral coupling.

\section*{Discussion}

We propose that the hierarchy of grid modules in the MEC is self-organized by competition in attractor networks between excitation along the longitudinal MEC axis and lateral inhibition. We showed that such an architecture, with an inhibition length scale that increases smoothly along the MEC axis, reproduces a central experimental finding: grid cells form modules with scales clustered around discrete values~\cite{Stensola:2012gn,Barry:2007gv,Krupic:2015gj}.  

The distribution of scales across modules in our model quantitatively matches experiments. Different groups have reported mean scale ratios of 1.64 (6 module pairs), 1.42 (24 module pairs), and 1.56 (11 module pairs)~\cite{Barry:2007gv,Stensola:2012gn,Krupic:2015gj}. These data could be interpreted as an indication that the grid system has a preferred scale ratio roughly in range of 1.4--1.7.  As we showed, our model naturally produces a hierarchy of modules with scale ratios in this range; its network parameters lead to both commensurate and discommensurate grids (\textbf{Fig.~\ref{fig:phase}}). On the other hand, the data on scale ratios between individual pairs of modules actually span a range of values in the different experiments: 1.6--1.9, 1.1--1.8, and 1.2--2.0~\cite{Barry:2007gv,Stensola:2012gn,Krupic:2015gj}. This suggests that the underlying mechanism that produces grid modules must be capable of producing different scale ratios as its parameters vary. This is indeed the case for our model, in which variation of network parameters produces a realistic range of scale ratios (\textbf{Fig.~\ref{fig:range}}). Despite variability across individual scale ratios, experiments strikingly reveal that the average scale ratio is the same from the smallest pair of modules to the largest pair, whereas the average scale \emph{difference} changes across the hierarchy~\cite{Stensola:2012gn}.  Our model robustly reproduces this observation (\textbf{Fig.~\ref{fig:range}c},~\textbf{d}) because its fundamental mechanism of geometric coordination between grids enforces constant-on-average scale ratios even with variation in parameters among individual networks.

Our model requires that grid orientation be co-modular with scale, as observed in experiment~\cite{Stensola:2012gn}.   Studies characterizing the statistics of orientation differences between modules are limited, but values seem to span the entire range \SI{0}{\degree}--\SI{30}{\degree}, with some preference for values at the low and high ends of this range~\cite{Krupic:2015gj}.  Our model can capture the entire range of orientation differences with discommensurate relationships favoring small differences and commensurate relationships favoring large differences (\textbf{Fig.~\ref{fig:phase}}). Overall, our model predicts a positive correlation between scale ratio and orientation difference (\textbf{Figs.~\ref{fig:phase}e} and \textbf{\ref{fig:range}b}), which can be tested experimentally.  Existing datasets~\cite{Stensola:2012gn,Krupic:2015gj} have a confound---animals are tested in square and rectangular enclosures which have distinguishable orientations marked by the corners. Grid orientations can anchor to such features~\cite{Stensola:2015cj}, either through the integration of visual and external cues~\cite{raudies2015differences,savelli2017framing}, or through interaction with boundaries~\cite{Bush:2014iq,krupic2016framing,giocomo2016environmental,Evans:2016cf,Hardcastle:2017ce,Keinath:2018el,Ocko:2018ed}. Experiments in circular or other non-rectangular environments may help disambiguate the effects of such anchoring. Our model also predicts that orientation differences between modules will be preserved between environments with different geometries since the differences are internally generated by the dynamics of the network. This effect has been observed~\cite{Krupic:2015gj}.

Our model produces consistent differences in firing rate from one grid field to another for some grid cells. This variability arises at module interfaces from the selective excitation of some network activity peaks in the smaller-scale grid by the overlapping activity peaks of the larger-scale grid. Such an explanation for firing rate variability is suggested by Ref.~\onlinecite{Ismakov:2017jj} and would be further supported by observing spatial periodicity in the variability corresponding to the scale of the larger grids. An alternative model, in which field-to-field firing rate variability arises from place cell inputs~\cite{Dunn:2017jk}, would not lead to such periodicity.

Our model requires excitatory coupling between grid cells at different locations along the longitudinal MEC axis, either through direct excitation or disinhibition~\cite{Fuchs:2016et}. As a result, it predicts that destruction of grid cells, or inactivation of excitatory coupling~\cite{Zutshi:2018ku}, at a given location along the axis will change grid scales and/or orientations at other locations (\textbf{Fig.~\ref{fig:lesion}}). The presence of noise correlations across modules, as previously investigated but not fully characterized~\cite{mathis2013multiscale,Tocker:2015ff}, would suggest connections between modules. Such correlations, and perhaps even lattice relationships, could be observed via calcium imaging of the MEC~\cite{Heys:2014cv,Gu:2018dm}. A direct test for coupling would involve patch clamp experiments akin to those used to identify local inhibition and excitation and interhemispheric excitation between principal cells in superficial MEC layers~\cite{Couey:2013fi,Fuchs:2016et,Winterer:2017fl}. 

Since spatial grid scales are both proportional to inhibition length scale $l$ and inversely proportional to velocity gain $\alpha$ (Ref.~\onlinecite{Burak:2009fx} and \textbf{Methods}), we also simulated excitatorily coupled networks with a depth-dependent velocity gain $\alpha(z)$ and a fixed inhibition distance $l$ (\textbf{Supp.\@ Info.}). In contrast to simulations in one dimension~\cite{widloskifiete}, while we observed module self-organization, the system gave inconsistent results among replicate simulations and lacked fixed scale ratios (\textbf{Supp.\@ Figs.~7} and \textbf{8} and \textbf{Supp.\@ Video}). Moreover, recent calcium imaging experiments suggest that activity on the MEC is arranged a deformed triangular lattice~\cite{Gu:2018dm}, as predicted by the continuous attractor model~\cite{Burak:2009fx}, and that regions with activity separated by larger anatomic distances contain grid cells of larger spatial scale. These observations support a changing inhibition length scale over a changing velocity gain as a mechanism for producing different grid scales, under the assumption that anatomic and network distances correspond to each other.

Our results differ from previous work on mechanisms for forming grid modules. Grossberg and Pilly hypothesize that grid cells arise from stripe cells in parasubiculum, and that discreteness in the spatial period of stripe cells leads to modularity of grid cells~\cite{Grossberg:2012ih}. However, stripe cells have only been observed once~\cite{Krupic:2012id,Navratilova:2016iq}, and the origin of discrete periods with constant-on-average ratios in stripe cells would then need to be addressed. Urdapilleta, Si, and Treves propose a model in which discrete modules self-organize from smooth gradients in parameters in a model where grid formation is driven by firing rate adaptation in single cells~\cite{Urdapilleta:2017kn}. They also utilize excitatory coupling among grid cells along the longitudinal MEC axis. However, this model does not have a mechanism to dynamically enforce the average constancy of grid scale ratios, which is a feature of the grid system~\cite{Stensola:2012gn}. The model also does not demonstrate modules with orientation differences near \SI{30}{\degree}~\cite{Krupic:2015gj}. Our model naturally reproduces these features of the grid system. Furthermore, over the past few years, multiple reports have provided independent experimental support for the importance of recurrent connections among grid cells~\cite{Couey:2013fi,Dunn:2015he,Fuchs:2016et,Zutshi:2018ku} and for the continuous attractor model in particular~\cite{Yoon:2013hv,Heys:2014cv,Gu:2018dm}. Our work establishes that continuous attractor networks can produce a discrete hierarchy of modules with a constant-on-average scale ratio.

The competition generated between excitatory and inhibitory connections bears a strong resemblance to the Frenkel-Kontorova model of condensed matter physics, in which a periodic potential of one scale acts on particles that prefer to form a lattice of a different, competing scale~\cite{Kontorova:1938en}. This model has a rich literature with many deep theoretical results, including the calculation of complicated phase diagrams involving ``devil's staircases''~\cite{Bak:1982it,chaikinlubensky} which mirror those of our model (\textbf{Fig.~\ref{fig:phase}}). Under certain conditions, our model produces networks with quasicrystalline approximant grids that are driven by networks with standard triangular grids at other scales (\textbf{Supp.\@ Fig.~9}). Quasicrystalline order lacks periodicity, but contains more nuanced positional order~\cite{Levine:1986kb}. This phenomenon wherein quasicrystalline structure is driven by crystalline order in a coupled system was recently observed for the first time in thin-film materials that contain Frenkel-Kontorova-like interactions~\cite{Forster:2013de,Forster:2016bd,Passens:2017kl}.

Commensurate and discommensurate lattice relationships are a robust and versatile mechanism for self-organizing a grid system whose scale ratios are constant or constant on average across a hierarchy of modules. We demonstrated this mechanism in a basic extension of the continuous attractor model with excitatory connections between networks. This model is amenable to extensions that capture other features of the grid system, such as spiking dynamics, learning of synaptic weights~\cite{Widloski:2014dl}, the union of our separate networks into a single network spanning the entire MEC, and the addition of border cell inputs or recurrent coupling between modules to correct path-integration errors or react to environmental deformations~\cite{Hardcastle:2015il,Keinath:2018el,Ocko:2018ed,pollock,mosheiffburak}.

\begin{acknowledgments}
  We are grateful to Xue-Xin Wei, Tom Lubensky, Ila Fiete, and John Widloski for their thoughtful ideas and suggestions, and to the Honda Research Institute and the NSF (grant PHY-1734030) for research support. L.K. is also supported by the Miller Institute for Basic Research in Science. Work on this project at the Aspen Center for Physics was supported by NSF grant PHY-1607611.
\end{acknowledgments}

\section*{Methods}

\begin{table}
	\caption{\label{tab:params}Main model parameters and their values unless otherwise noted.}
	\begin{ruledtabular}\begin{tabular}{ccc}
		Parameter & Variable & Value \\
		\hline
		Number of networks & $h$ & 12 \\
		Number of neurons per network & $n \times n$ & $160 \times 160$ \\
		Neurons recorded per network & & $3$ \\
		Animal speed & $|\ve V|$ & 0--\SI{1}{\m\per\s} \\
		Diameter of enclosure & & \SI{180}{\cm} \\
		Simulation time & & \SI{500}{\s} \\
		Simulation timestep & $\Delta t$ & \SI{1}{\ms} \\
		Neural relaxation time & $\tau$ & \SI{10}{\ms} \\
		Hippocampal input strength & $a_\textrm{mag}$ & 1 \\
		Hippocampal input falloff & $a_\textrm{fall}$ & 4 \\
		Inhibition distance minimum & $l_\textrm{min}$ & 4 \\
		Inhibition distance maximum & $l_\textrm{max}$ & 15 \\
		Inhibition distance exponent & $l_\textrm{exp}$ & $-1$ \\
		Inhibition strength & $w_\textrm{mag}$ & 2.4 \\
		Subpopulation shift & $\xi$ & 1 \\
		Coupling spread & $d$ & 8 \\
		Coupling strength & $u_\textrm{mag}$ & 2.6 \\
		Velocity gain & $\alpha$ & \SI{0.3}{\s\per\m}
	\end{tabular}\end{ruledtabular}
\end{table}

\subsection*{Model setup and dynamics}

We implemented the Burak-Fiete model as follows~\cite{Burak:2009fx}. Networks $z = 1, \ldots, h$ each contain a 2D sheet of neurons with indices $\ve r = (x, y)$, where $x = 1, \ldots, n$ and $y = 1, \ldots, n$. Neurons receive broad excitatory input $a(\ve r)$ from the hippocampus, and, to prevent edge effects, those toward the center of the networks receive more excitation than those toward the edges. Each neuron also inhibits others that lie around a length scale of $l(z)$ neurons away in the same network $z$  Moreover, every neuron belongs to one of four subpopulations that evenly tile the neural sheet. Each subpopulation is associated with both a preferred direction $\veh e$ along one of the network axes $\pm \veh x$ or $\pm \veh y$ and a corresponding preferred direction $\veh E$ along an axis $\pm \veh X$ or $\pm \veh Y$ in its spatial environment. A neuron at position $\ve r$ in network $z$ has its inhibitory outputs $w(\ve r; z)$ shifted slightly by $\xi$ neurons in the $\veh e(\ve r)$ direction and its hippocampal excitation modulated by a small amount proportional to $\veh E(\ve r) \cdot \ve V$, where $\ve V$ is the spatial velocity of the animal.  Note that lowercase letters refer to attractor networks at each depth $z$ in which distances have units of neurons, and uppercase letters refer to the animal's spatial environment in which distances have physical units, such as centimeters.

In addition to these established features~\cite{Burak:2009fx}, we introduce excitatory connections $u(\ve r)$ from every neuron $\ve r$ in network $z$ to neurons located within a spread $d$ of the same $\ve r$ but in the preceding network with depth $z-1$. $u(\ve r)$ is constant for all networks except for the last one $z = h$, which has $u(\ve r) = 0$. These components lead to the following dynamical equation for the dimensionless neural firing rates $s(\ve r, z, t)$:
\begin{eqnarray}
	&&\tau \frac{s(\ve r,z,t+\Delta t)-s(\ve r,z,t)}{\Delta t} + s(\ve r,z,t) \nonum
	&& \quad{}= \bigg\{\sum_{\ve r'} w(\ve r-\ve r'+\xi \veh e(\ve r');z) s(\ve r',z,t) \nonum
	&& \quad\qquad{}+ \sum_{\ve r'} u(\ve r-\ve r') s(\ve r',z+1,t) \nonum
	&& \quad\qquad{}+ a(\ve r) \left[1 + \alpha \veh E(\ve r) \cdot \ve V(t)\right]\bigg\}_+.
	\label{eqn:s}
\end{eqnarray}
Inputs to each neuron are rectified by $\{c\}_+ = 0$ for $c < 0$, $c$ for $c \geq 0$. $\Delta t$ is the simulation time increment, $\tau$ is the neural relaxation time, and $\alpha$ is the velocity gain that describes how much the animal's velocity $\ve V$ modulates the hippocampal inputs $a(\ve r)$. Note that $s$ can be treated as a dimensionless variable because \textbf{Eq.~\ref{eqn:s}} is invariant to scaling of $s$ and $a$ by the same factor.

We use velocities $\ve V(t)$ corresponding to a real rat trajectory~\cite{Hafting:2005dp, Burak:2009fx}. Details are provided in \textbf{Supp.\@ Info.}

\subsection*{Inhibitory and excitatory connections}

The hippocampal input is
\begin{equation}
	a(\ve r) = \begin{cases} a_\textrm{mag} \ee^{-a_\textrm{fall} r_\textrm{scaled}^2} & r_\textrm{scaled} < 1 \\ 0 & r_\textrm{scaled} \geq 1, \end{cases}
	\label{eqn:a}
\end{equation}
where $r_\textrm{scaled} = \sqrt{\left(x-\frac{n+1}{2}\right)^2 + \left(y-\frac{n+1}{2}\right)^2}/\frac{n}{2}$ is a scaled radial distance for the neuron at $\ve r = (x, y)$, $a_\textrm{mag}$ is the magnitude of the input, and $a_\textrm{fall}$ is a falloff parameter. The inhibition distance for network $z$ is
\begin{equation}
	l(z) = \left[l_\textrm{min}^{l_\textrm{exp}} + \left(l_\textrm{max}^{l_\textrm{exp}} - l_\textrm{min}^{l_\textrm{exp}}\right) \frac{z-1}{h-1}\right]^{1/l_\textrm{exp}},
	\label{eqn:l}
\end{equation}
which ranges from $l_\textrm{min} = l(1)$ to $l_\textrm{max} = l(h)$ with concavity tuned by $l_\textrm{exp}$. More negative values of $l_\textrm{exp}$ lead to greater concavity; for $l_\textrm{exp} = 0$, we use the limiting expression $l(z) = l_\textrm{min}^{(h-z)/(h-1)} l_\textrm{max}^{(z-1)/(h-1)}$.
The recurrent inhibition profile for network $z$ is
\begin{equation}
	w(\ve r; z) = \begin{cases} -\dfrac{w_\textrm{mag}}{l(z)^2} \dfrac{1-\cos[\pi r/l(z)]}{2} & r < 2 l(z) \\ 0 & r \geq 2 l(z), \end{cases}
	\label{eqn:w}
\end{equation}
where $w_\textrm{mag}$ is the magnitude of inhibition. We scale this magnitude by $l(z)^{-2}$ to make the integrated inhibition constant across $z$. The excitatory coupling is
\begin{equation}
	u(\ve r) = \begin{cases} \dfrac{u_\textrm{mag}}{d^2} \dfrac{1+\cos[\pi r/d]}{2} & r < d \\ 0 & r \geq d, \end{cases}
	\label{eqn:u}
\end{equation}
where $u_\textrm{mag}$ and $d$ are the magnitude and spread of coupling, respectively. In analogy to $w_\textrm{mag}$, we scale $u_\textrm{mag}$ by $d^{-2}$.

\subsection*{Overview of data analysis techniques}
To determine spatial grid scales, orientations, and gridness, we consider an annular region of the spatial autocorrelation map that contains the 6 peaks closest to the origin. Grid scale is the radius with highest value, averaging over angles. Grid orientation and gridness are determined by first averaging over radial distance and analyzing the sixth component of the Fourier series with respect to angle~\cite{Weber:2019gb}.  The power of this component divided by the total Fourier power measures ``gridness'' and its complex phase measures the orientation. Grid cells are subject to a gridness cutoff of 0.6. For each replicate simulation, we cluster its grid cells with respect to scale and orientation using a $k$-means procedure with $k$ determined by kernel smoothed densities~\cite{Stensola:2012gn}. See \textbf{Supp.\@ Info.} for full details.


\begin{thebibliography}{61}%
\makeatletter
\providecommand \@ifxundefined [1]{%
 \@ifx{#1\undefined}
}%
\providecommand \@ifnum [1]{%
 \ifnum #1\expandafter \@firstoftwo
 \else \expandafter \@secondoftwo
 \fi
}%
\providecommand \@ifx [1]{%
 \ifx #1\expandafter \@firstoftwo
 \else \expandafter \@secondoftwo
 \fi
}%
\providecommand \natexlab [1]{#1}%
\providecommand \enquote  [1]{``#1''}%
\providecommand \bibnamefont  [1]{#1}%
\providecommand \bibfnamefont [1]{#1}%
\providecommand \citenamefont [1]{#1}%
\providecommand \href@noop [0]{\@secondoftwo}%
\providecommand \href [0]{\begingroup \@sanitize@url \@href}%
\providecommand \@href[1]{\@@startlink{#1}\@@href}%
\providecommand \@@href[1]{\endgroup#1\@@endlink}%
\providecommand \@sanitize@url [0]{\catcode `\\12\catcode `\$12\catcode
  `\&12\catcode `\#12\catcode `\^12\catcode `\_12\catcode `\%12\relax}%
\providecommand \@@startlink[1]{}%
\providecommand \@@endlink[0]{}%
\providecommand \url  [0]{\begingroup\@sanitize@url \@url }%
\providecommand \@url [1]{\endgroup\@href {#1}{\urlprefix }}%
\providecommand \urlprefix  [0]{URL }%
\providecommand \Eprint [0]{\href }%
\providecommand \doibase [0]{http://dx.doi.org/}%
\providecommand \selectlanguage [0]{\@gobble}%
\providecommand \bibinfo  [0]{\@secondoftwo}%
\providecommand \bibfield  [0]{\@secondoftwo}%
\providecommand \translation [1]{[#1]}%
\providecommand \BibitemOpen [0]{}%
\providecommand \bibitemStop [0]{}%
\providecommand \bibitemNoStop [0]{.\EOS\space}%
\providecommand \EOS [0]{\spacefactor3000\relax}%
\providecommand \BibitemShut  [1]{\csname bibitem#1\endcsname}%
\let\auto@bib@innerbib\@empty
\bibitem [{\citenamefont {Hafting}\ \emph {et~al.}(2005)\citenamefont
  {Hafting}, \citenamefont {Fyhn}, \citenamefont {Molden}, \citenamefont
  {Moser},\ and\ \citenamefont {Moser}}]{Hafting:2005dp}%
  \BibitemOpen
  \bibfield  {author} {\bibinfo {author} {\bibfnamefont {T.}~\bibnamefont
  {Hafting}}, \bibinfo {author} {\bibfnamefont {M.}~\bibnamefont {Fyhn}},
  \bibinfo {author} {\bibfnamefont {S.}~\bibnamefont {Molden}}, \bibinfo
  {author} {\bibfnamefont {M.-B.}\ \bibnamefont {Moser}}, \ and\ \bibinfo
  {author} {\bibfnamefont {E.~I.}\ \bibnamefont {Moser}},\ }\href@noop {}
  {\bibfield  {journal} {\bibinfo  {journal} {Nature}\ }\textbf {\bibinfo
  {volume} {436}},\ \bibinfo {pages} {801} (\bibinfo {year}
  {2005})}\BibitemShut {NoStop}%
\bibitem [{\citenamefont {Stensola}\ \emph {et~al.}(2012)\citenamefont
  {Stensola}, \citenamefont {Stensola}, \citenamefont {Solstad}, \citenamefont
  {Fr{\o}land}, \citenamefont {Moser},\ and\ \citenamefont
  {Moser}}]{Stensola:2012gn}%
  \BibitemOpen
  \bibfield  {author} {\bibinfo {author} {\bibfnamefont {H.}~\bibnamefont
  {Stensola}}, \bibinfo {author} {\bibfnamefont {T.}~\bibnamefont {Stensola}},
  \bibinfo {author} {\bibfnamefont {T.}~\bibnamefont {Solstad}}, \bibinfo
  {author} {\bibfnamefont {K.}~\bibnamefont {Fr{\o}land}}, \bibinfo {author}
  {\bibfnamefont {M.-B.}\ \bibnamefont {Moser}}, \ and\ \bibinfo {author}
  {\bibfnamefont {E.~I.}\ \bibnamefont {Moser}},\ }\href@noop {} {\bibfield
  {journal} {\bibinfo  {journal} {Nature}\ }\textbf {\bibinfo {volume} {492}},\
  \bibinfo {pages} {72} (\bibinfo {year} {2012})}\BibitemShut {NoStop}%
\bibitem [{\citenamefont {Barry}\ \emph {et~al.}(2007)\citenamefont {Barry},
  \citenamefont {Hayman}, \citenamefont {Burgess},\ and\ \citenamefont
  {Jeffery}}]{Barry:2007gv}%
  \BibitemOpen
  \bibfield  {author} {\bibinfo {author} {\bibfnamefont {C.}~\bibnamefont
  {Barry}}, \bibinfo {author} {\bibfnamefont {R.}~\bibnamefont {Hayman}},
  \bibinfo {author} {\bibfnamefont {N.}~\bibnamefont {Burgess}}, \ and\
  \bibinfo {author} {\bibfnamefont {K.~J.}\ \bibnamefont {Jeffery}},\
  }\href@noop {} {\bibfield  {journal} {\bibinfo  {journal} {Nat. Neurosci.}\
  }\textbf {\bibinfo {volume} {10}},\ \bibinfo {pages} {682} (\bibinfo {year}
  {2007})}\BibitemShut {NoStop}%
\bibitem [{\citenamefont {Krupic}\ \emph {et~al.}(2015)\citenamefont {Krupic},
  \citenamefont {Bauza}, \citenamefont {Burton}, \citenamefont {Barry},\ and\
  \citenamefont {O'Keefe}}]{Krupic:2015gj}%
  \BibitemOpen
  \bibfield  {author} {\bibinfo {author} {\bibfnamefont {J.}~\bibnamefont
  {Krupic}}, \bibinfo {author} {\bibfnamefont {M.}~\bibnamefont {Bauza}},
  \bibinfo {author} {\bibfnamefont {S.}~\bibnamefont {Burton}}, \bibinfo
  {author} {\bibfnamefont {C.}~\bibnamefont {Barry}}, \ and\ \bibinfo {author}
  {\bibfnamefont {J.}~\bibnamefont {O'Keefe}},\ }\href@noop {} {\bibfield
  {journal} {\bibinfo  {journal} {Nature}\ }\textbf {\bibinfo {volume} {518}},\
  \bibinfo {pages} {232} (\bibinfo {year} {2015})}\BibitemShut {NoStop}%
\bibitem [{\citenamefont {Moser}\ \emph {et~al.}(2008)\citenamefont {Moser},
  \citenamefont {Kropff},\ and\ \citenamefont {Moser}}]{Moser:2008hh}%
  \BibitemOpen
  \bibfield  {author} {\bibinfo {author} {\bibfnamefont {E.~I.}\ \bibnamefont
  {Moser}}, \bibinfo {author} {\bibfnamefont {E.}~\bibnamefont {Kropff}}, \
  and\ \bibinfo {author} {\bibfnamefont {M.-B.}\ \bibnamefont {Moser}},\
  }\href@noop {} {\bibfield  {journal} {\bibinfo  {journal} {Annu. Rev.
  Neurosci.}\ }\textbf {\bibinfo {volume} {31}},\ \bibinfo {pages} {69}
  (\bibinfo {year} {2008})}\BibitemShut {NoStop}%
\bibitem [{\citenamefont {Fiete}\ \emph {et~al.}(2008)\citenamefont {Fiete},
  \citenamefont {Burak},\ and\ \citenamefont {Brookings}}]{Fiete:2008dz}%
  \BibitemOpen
  \bibfield  {author} {\bibinfo {author} {\bibfnamefont {I.~R.}\ \bibnamefont
  {Fiete}}, \bibinfo {author} {\bibfnamefont {Y.}~\bibnamefont {Burak}}, \ and\
  \bibinfo {author} {\bibfnamefont {T.}~\bibnamefont {Brookings}},\ }\href@noop
  {} {\bibfield  {journal} {\bibinfo  {journal} {J. Neurosci.}\ }\textbf
  {\bibinfo {volume} {28}},\ \bibinfo {pages} {6858} (\bibinfo {year}
  {2008})}\BibitemShut {NoStop}%
\bibitem [{\citenamefont {Mathis}\ \emph {et~al.}(2012)\citenamefont {Mathis},
  \citenamefont {Herz},\ and\ \citenamefont {Stemmler}}]{mathis2012optimal}%
  \BibitemOpen
  \bibfield  {author} {\bibinfo {author} {\bibfnamefont {A.}~\bibnamefont
  {Mathis}}, \bibinfo {author} {\bibfnamefont {A.~V.~M.}\ \bibnamefont {Herz}},
  \ and\ \bibinfo {author} {\bibfnamefont {M.}~\bibnamefont {Stemmler}},\
  }\href@noop {} {\bibfield  {journal} {\bibinfo  {journal} {Neural
  Computation}\ }\textbf {\bibinfo {volume} {24}},\ \bibinfo {pages} {2280}
  (\bibinfo {year} {2012})}\BibitemShut {NoStop}%
\bibitem [{\citenamefont {Wei}\ \emph {et~al.}(2015)\citenamefont {Wei},
  \citenamefont {Prentice},\ and\ \citenamefont
  {Balasubramanian}}]{Wei:2015hl}%
  \BibitemOpen
  \bibfield  {author} {\bibinfo {author} {\bibfnamefont {X.-X.}\ \bibnamefont
  {Wei}}, \bibinfo {author} {\bibfnamefont {J.}~\bibnamefont {Prentice}}, \
  and\ \bibinfo {author} {\bibfnamefont {V.}~\bibnamefont {Balasubramanian}},\
  }\href@noop {} {\bibfield  {journal} {\bibinfo  {journal} {eLife}\ }
  (\bibinfo {year} {2015})}\BibitemShut {NoStop}%
\bibitem [{\citenamefont {Stemmler}\ \emph {et~al.}(2015)\citenamefont
  {Stemmler}, \citenamefont {Mathis},\ and\ \citenamefont
  {Herz}}]{Stemmler:2015gc}%
  \BibitemOpen
  \bibfield  {author} {\bibinfo {author} {\bibfnamefont {M.}~\bibnamefont
  {Stemmler}}, \bibinfo {author} {\bibfnamefont {A.}~\bibnamefont {Mathis}}, \
  and\ \bibinfo {author} {\bibfnamefont {A.~V.~M.}\ \bibnamefont {Herz}},\
  }\href@noop {} {\bibfield  {journal} {\bibinfo  {journal} {Science Advances}\
  }\textbf {\bibinfo {volume} {1}},\ \bibinfo {pages} {e1500816} (\bibinfo
  {year} {2015})}\BibitemShut {NoStop}%
\bibitem [{\citenamefont {Sanzeni}\ \emph {et~al.}(2016)\citenamefont
  {Sanzeni}, \citenamefont {Balasubramanian}, \citenamefont {Tiana},\ and\
  \citenamefont {Vergassola}}]{Sanzeni:2016fg}%
  \BibitemOpen
  \bibfield  {author} {\bibinfo {author} {\bibfnamefont {A.}~\bibnamefont
  {Sanzeni}}, \bibinfo {author} {\bibfnamefont {V.}~\bibnamefont
  {Balasubramanian}}, \bibinfo {author} {\bibfnamefont {G.}~\bibnamefont
  {Tiana}}, \ and\ \bibinfo {author} {\bibfnamefont {M.}~\bibnamefont
  {Vergassola}},\ }\href@noop {} {\bibfield  {journal} {\bibinfo  {journal}
  {Phys. Rev. E}\ }\textbf {\bibinfo {volume} {94}},\ \bibinfo {pages} {599}
  (\bibinfo {year} {2016})}\BibitemShut {NoStop}%
\bibitem [{\citenamefont {Mosheiff}\ \emph {et~al.}(2017)\citenamefont
  {Mosheiff}, \citenamefont {Agmon}, \citenamefont {Moriel},\ and\
  \citenamefont {Burak}}]{Mosheiff:2017fj}%
  \BibitemOpen
  \bibfield  {author} {\bibinfo {author} {\bibfnamefont {N.}~\bibnamefont
  {Mosheiff}}, \bibinfo {author} {\bibfnamefont {H.}~\bibnamefont {Agmon}},
  \bibinfo {author} {\bibfnamefont {A.}~\bibnamefont {Moriel}}, \ and\ \bibinfo
  {author} {\bibfnamefont {Y.}~\bibnamefont {Burak}},\ }\href@noop {}
  {\bibfield  {journal} {\bibinfo  {journal} {PLOS Comp. Biol.}\ }\textbf
  {\bibinfo {volume} {13}},\ \bibinfo {pages} {e1005597} (\bibinfo {year}
  {2017})}\BibitemShut {NoStop}%
\bibitem [{\citenamefont {Sreenivasan}\ and\ \citenamefont
  {Fiete}(2011)}]{Sreenivasan:2011fy}%
  \BibitemOpen
  \bibfield  {author} {\bibinfo {author} {\bibfnamefont {S.}~\bibnamefont
  {Sreenivasan}}\ and\ \bibinfo {author} {\bibfnamefont {I.}~\bibnamefont
  {Fiete}},\ }\href@noop {} {\bibfield  {journal} {\bibinfo  {journal} {Nat.
  Neurosci.}\ }\textbf {\bibinfo {volume} {14}},\ \bibinfo {pages} {1330}
  (\bibinfo {year} {2011})}\BibitemShut {NoStop}%
\bibitem [{\citenamefont {Cueva}\ and\ \citenamefont
  {Wei}(2018)}]{cueva2018emergence}%
  \BibitemOpen
  \bibfield  {author} {\bibinfo {author} {\bibfnamefont {C.~J.}\ \bibnamefont
  {Cueva}}\ and\ \bibinfo {author} {\bibfnamefont {X.-X.}\ \bibnamefont
  {Wei}},\ }\href@noop {} {\bibfield  {journal} {\bibinfo  {journal}
  {International Conference on Learning Representations}\ } (\bibinfo {year}
  {2018})}\BibitemShut {NoStop}%
\bibitem [{\citenamefont {Banino}\ \emph {et~al.}(2018)\citenamefont {Banino},
  \citenamefont {Barry}, \citenamefont {Uria}, \citenamefont {Blundell},
  \citenamefont {Lillicrap}, \citenamefont {Mirowski}, \citenamefont {Pritzel},
  \citenamefont {Chadwick}, \citenamefont {Degris}, \citenamefont {Modayil}
  \emph {et~al.}}]{banino2018vector}%
  \BibitemOpen
  \bibfield  {author} {\bibinfo {author} {\bibfnamefont {A.}~\bibnamefont
  {Banino}}, \bibinfo {author} {\bibfnamefont {C.}~\bibnamefont {Barry}},
  \bibinfo {author} {\bibfnamefont {B.}~\bibnamefont {Uria}}, \bibinfo {author}
  {\bibfnamefont {C.}~\bibnamefont {Blundell}}, \bibinfo {author}
  {\bibfnamefont {T.}~\bibnamefont {Lillicrap}}, \bibinfo {author}
  {\bibfnamefont {P.}~\bibnamefont {Mirowski}}, \bibinfo {author}
  {\bibfnamefont {A.}~\bibnamefont {Pritzel}}, \bibinfo {author} {\bibfnamefont
  {M.~J.}\ \bibnamefont {Chadwick}}, \bibinfo {author} {\bibfnamefont
  {T.}~\bibnamefont {Degris}}, \bibinfo {author} {\bibfnamefont
  {J.}~\bibnamefont {Modayil}},  \emph {et~al.},\ }\href@noop {} {\bibfield
  {journal} {\bibinfo  {journal} {Nature}\ ,\ \bibinfo {pages} {1}} (\bibinfo
  {year} {2018})}\BibitemShut {NoStop}%
\bibitem [{\citenamefont {Fuhs}\ and\ \citenamefont
  {Touretzky}(2006)}]{Fuhs:2006fb}%
  \BibitemOpen
  \bibfield  {author} {\bibinfo {author} {\bibfnamefont {M.~C.}\ \bibnamefont
  {Fuhs}}\ and\ \bibinfo {author} {\bibfnamefont {D.~S.}\ \bibnamefont
  {Touretzky}},\ }\href@noop {} {\bibfield  {journal} {\bibinfo  {journal} {J.
  Neurosci.}\ }\textbf {\bibinfo {volume} {26}},\ \bibinfo {pages} {4266}
  (\bibinfo {year} {2006})}\BibitemShut {NoStop}%
\bibitem [{\citenamefont {Burak}\ and\ \citenamefont
  {Fiete}(2009)}]{Burak:2009fx}%
  \BibitemOpen
  \bibfield  {author} {\bibinfo {author} {\bibfnamefont {Y.}~\bibnamefont
  {Burak}}\ and\ \bibinfo {author} {\bibfnamefont {I.~R.}\ \bibnamefont
  {Fiete}},\ }\href@noop {} {\bibfield  {journal} {\bibinfo  {journal} {PLOS
  Comp. Biol.}\ }\textbf {\bibinfo {volume} {5}},\ \bibinfo {pages} {e1000291}
  (\bibinfo {year} {2009})}\BibitemShut {NoStop}%
\bibitem [{\citenamefont {Burgess}\ \emph {et~al.}(2007)\citenamefont
  {Burgess}, \citenamefont {Barry},\ and\ \citenamefont
  {O'Keefe}}]{Burgess:2007fi}%
  \BibitemOpen
  \bibfield  {author} {\bibinfo {author} {\bibfnamefont {N.}~\bibnamefont
  {Burgess}}, \bibinfo {author} {\bibfnamefont {C.}~\bibnamefont {Barry}}, \
  and\ \bibinfo {author} {\bibfnamefont {J.}~\bibnamefont {O'Keefe}},\
  }\href@noop {} {\bibfield  {journal} {\bibinfo  {journal} {Hippocampus}\
  }\textbf {\bibinfo {volume} {17}},\ \bibinfo {pages} {801} (\bibinfo {year}
  {2007})}\BibitemShut {NoStop}%
\bibitem [{\citenamefont {Hasselmo}\ \emph {et~al.}(2007)\citenamefont
  {Hasselmo}, \citenamefont {Giocomo},\ and\ \citenamefont
  {Zilli}}]{Hasselmo:2007cv}%
  \BibitemOpen
  \bibfield  {author} {\bibinfo {author} {\bibfnamefont {M.~E.}\ \bibnamefont
  {Hasselmo}}, \bibinfo {author} {\bibfnamefont {L.~M.}\ \bibnamefont
  {Giocomo}}, \ and\ \bibinfo {author} {\bibfnamefont {E.~A.}\ \bibnamefont
  {Zilli}},\ }\href@noop {} {\bibfield  {journal} {\bibinfo  {journal}
  {Hippocampus}\ }\textbf {\bibinfo {volume} {17}},\ \bibinfo {pages} {1252}
  (\bibinfo {year} {2007})}\BibitemShut {NoStop}%
\bibitem [{\citenamefont {Bush}\ and\ \citenamefont
  {Burgess}(2014)}]{Bush:2014iq}%
  \BibitemOpen
  \bibfield  {author} {\bibinfo {author} {\bibfnamefont {D.}~\bibnamefont
  {Bush}}\ and\ \bibinfo {author} {\bibfnamefont {N.}~\bibnamefont {Burgess}},\
  }\href@noop {} {\bibfield  {journal} {\bibinfo  {journal} {J. Neurosci.}\
  }\textbf {\bibinfo {volume} {34}},\ \bibinfo {pages} {5065} (\bibinfo {year}
  {2014})}\BibitemShut {NoStop}%
\bibitem [{\citenamefont {Issa}\ \emph {et~al.}(2008)\citenamefont {Issa},
  \citenamefont {Rosenberg},\ and\ \citenamefont {Husson}}]{Issa:2008de}%
  \BibitemOpen
  \bibfield  {author} {\bibinfo {author} {\bibfnamefont {N.~P.}\ \bibnamefont
  {Issa}}, \bibinfo {author} {\bibfnamefont {A.}~\bibnamefont {Rosenberg}}, \
  and\ \bibinfo {author} {\bibfnamefont {T.~R.}\ \bibnamefont {Husson}},\
  }\href@noop {} {\bibfield  {journal} {\bibinfo  {journal} {J. Neurophysiol.}\
  }\textbf {\bibinfo {volume} {99}},\ \bibinfo {pages} {2745} (\bibinfo {year}
  {2008})}\BibitemShut {NoStop}%
\bibitem [{\citenamefont {Taube}\ \emph {et~al.}(1990)\citenamefont {Taube},
  \citenamefont {Muller},\ and\ \citenamefont {Ranck}}]{Taube:1990vf}%
  \BibitemOpen
  \bibfield  {author} {\bibinfo {author} {\bibfnamefont {J.~S.}\ \bibnamefont
  {Taube}}, \bibinfo {author} {\bibfnamefont {R.~U.}\ \bibnamefont {Muller}}, \
  and\ \bibinfo {author} {\bibfnamefont {J.~B.}\ \bibnamefont {Ranck}},\
  }\href@noop {} {\bibfield  {journal} {\bibinfo  {journal} {J. Neurosci.}\
  }\textbf {\bibinfo {volume} {10}},\ \bibinfo {pages} {420} (\bibinfo {year}
  {1990})}\BibitemShut {NoStop}%
\bibitem [{\citenamefont {Grossberg}\ and\ \citenamefont
  {Pilly}(2012)}]{Grossberg:2012ih}%
  \BibitemOpen
  \bibfield  {author} {\bibinfo {author} {\bibfnamefont {S.}~\bibnamefont
  {Grossberg}}\ and\ \bibinfo {author} {\bibfnamefont {P.~K.}\ \bibnamefont
  {Pilly}},\ }\href@noop {} {\bibfield  {journal} {\bibinfo  {journal} {PLOS
  Comp. Biol.}\ }\textbf {\bibinfo {volume} {8}},\ \bibinfo {pages} {e1002648}
  (\bibinfo {year} {2012})}\BibitemShut {NoStop}%
\bibitem [{\citenamefont {Urdapilleta}\ \emph {et~al.}(2017)\citenamefont
  {Urdapilleta}, \citenamefont {Si},\ and\ \citenamefont
  {Treves}}]{Urdapilleta:2017kn}%
  \BibitemOpen
  \bibfield  {author} {\bibinfo {author} {\bibfnamefont {E.}~\bibnamefont
  {Urdapilleta}}, \bibinfo {author} {\bibfnamefont {B.}~\bibnamefont {Si}}, \
  and\ \bibinfo {author} {\bibfnamefont {A.}~\bibnamefont {Treves}},\
  }\href@noop {} {\bibfield  {journal} {\bibinfo  {journal} {Hippocampus}\
  }\textbf {\bibinfo {volume} {77}},\ \bibinfo {pages} {137} (\bibinfo {year}
  {2017})}\BibitemShut {NoStop}%
\bibitem [{\citenamefont {Bonnevie}\ \emph {et~al.}(2013)\citenamefont
  {Bonnevie}, \citenamefont {Dunn}, \citenamefont {Fyhn}, \citenamefont
  {Hafting}, \citenamefont {Derdikman}, \citenamefont {Kubie}, \citenamefont
  {Roudi}, \citenamefont {Moser},\ and\ \citenamefont
  {Moser}}]{Bonnevie:2013eu}%
  \BibitemOpen
  \bibfield  {author} {\bibinfo {author} {\bibfnamefont {T.}~\bibnamefont
  {Bonnevie}}, \bibinfo {author} {\bibfnamefont {B.}~\bibnamefont {Dunn}},
  \bibinfo {author} {\bibfnamefont {M.}~\bibnamefont {Fyhn}}, \bibinfo {author}
  {\bibfnamefont {T.}~\bibnamefont {Hafting}}, \bibinfo {author} {\bibfnamefont
  {D.}~\bibnamefont {Derdikman}}, \bibinfo {author} {\bibfnamefont {J.~L.}\
  \bibnamefont {Kubie}}, \bibinfo {author} {\bibfnamefont {Y.}~\bibnamefont
  {Roudi}}, \bibinfo {author} {\bibfnamefont {E.~I.}\ \bibnamefont {Moser}}, \
  and\ \bibinfo {author} {\bibfnamefont {M.-B.}\ \bibnamefont {Moser}},\
  }\href@noop {} {\bibfield  {journal} {\bibinfo  {journal} {Nat. Neurosci.}\
  }\textbf {\bibinfo {volume} {16}},\ \bibinfo {pages} {309} (\bibinfo {year}
  {2013})}\BibitemShut {NoStop}%
\bibitem [{\citenamefont {Chaikin}\ and\ \citenamefont
  {Lubensky}(1995)}]{chaikinlubensky}%
  \BibitemOpen
  \bibfield  {author} {\bibinfo {author} {\bibfnamefont {P.~M.}\ \bibnamefont
  {Chaikin}}\ and\ \bibinfo {author} {\bibfnamefont {T.~C.}\ \bibnamefont
  {Lubensky}},\ }\href@noop {} {\emph {\bibinfo {title} {Principles of
  Condensed Matter Physics}}}\ (\bibinfo  {publisher} {Cambridge University
  Press},\ \bibinfo {address} {Cambridge},\ \bibinfo {year} {1995})\ pp.\
  \bibinfo {pages} {601--620}\BibitemShut {NoStop}%
\bibitem [{\citenamefont {Ismakov}\ \emph {et~al.}(2017)\citenamefont
  {Ismakov}, \citenamefont {Barak}, \citenamefont {Jeffery},\ and\
  \citenamefont {Derdikman}}]{Ismakov:2017jj}%
  \BibitemOpen
  \bibfield  {author} {\bibinfo {author} {\bibfnamefont {R.}~\bibnamefont
  {Ismakov}}, \bibinfo {author} {\bibfnamefont {O.}~\bibnamefont {Barak}},
  \bibinfo {author} {\bibfnamefont {K.}~\bibnamefont {Jeffery}}, \ and\
  \bibinfo {author} {\bibfnamefont {D.}~\bibnamefont {Derdikman}},\ }\href@noop
  {} {\bibfield  {journal} {\bibinfo  {journal} {Curr. Biol.}\ }\textbf
  {\bibinfo {volume} {27}},\ \bibinfo {pages} {2337} (\bibinfo {year}
  {2017})}\BibitemShut {NoStop}%
\bibitem [{\citenamefont {Dunn}\ \emph {et~al.}(2017)\citenamefont {Dunn},
  \citenamefont {Wennberg}, \citenamefont {Huang},\ and\ \citenamefont
  {Roudi}}]{Dunn:2017jk}%
  \BibitemOpen
  \bibfield  {author} {\bibinfo {author} {\bibfnamefont {B.}~\bibnamefont
  {Dunn}}, \bibinfo {author} {\bibfnamefont {D.}~\bibnamefont {Wennberg}},
  \bibinfo {author} {\bibfnamefont {Z.}~\bibnamefont {Huang}}, \ and\ \bibinfo
  {author} {\bibfnamefont {Y.}~\bibnamefont {Roudi}},\ }\href@noop {}
  {\bibfield  {journal} {\bibinfo  {journal} {bioRxiv 10.1101/101899}\ }
  (\bibinfo {year} {2017})}\BibitemShut {NoStop}%
\bibitem [{\citenamefont {Wilson}(1990)}]{Wilson:1990cj}%
  \BibitemOpen
  \bibfield  {author} {\bibinfo {author} {\bibfnamefont {J.~A.}\ \bibnamefont
  {Wilson}},\ }\href@noop {} {\bibfield  {journal} {\bibinfo  {journal} {J.
  Phys.: Condens. Matter}\ }\textbf {\bibinfo {volume} {2}},\ \bibinfo {pages}
  {1683} (\bibinfo {year} {1990})}\BibitemShut {NoStop}%
\bibitem [{\citenamefont {Stensola}\ \emph {et~al.}(2015)\citenamefont
  {Stensola}, \citenamefont {Stensola}, \citenamefont {Moser},\ and\
  \citenamefont {Moser}}]{Stensola:2015cj}%
  \BibitemOpen
  \bibfield  {author} {\bibinfo {author} {\bibfnamefont {T.}~\bibnamefont
  {Stensola}}, \bibinfo {author} {\bibfnamefont {H.}~\bibnamefont {Stensola}},
  \bibinfo {author} {\bibfnamefont {M.-B.}\ \bibnamefont {Moser}}, \ and\
  \bibinfo {author} {\bibfnamefont {E.~I.}\ \bibnamefont {Moser}},\ }\href@noop
  {} {\bibfield  {journal} {\bibinfo  {journal} {Nature}\ }\textbf {\bibinfo
  {volume} {518}},\ \bibinfo {pages} {207} (\bibinfo {year}
  {2015})}\BibitemShut {NoStop}%
\bibitem [{\citenamefont {Raudies}\ and\ \citenamefont
  {Hasselmo}(2015)}]{raudies2015differences}%
  \BibitemOpen
  \bibfield  {author} {\bibinfo {author} {\bibfnamefont {F.}~\bibnamefont
  {Raudies}}\ and\ \bibinfo {author} {\bibfnamefont {M.~E.}\ \bibnamefont
  {Hasselmo}},\ }\href@noop {} {\bibfield  {journal} {\bibinfo  {journal} {PLOS
  Comp. Biol.}\ }\textbf {\bibinfo {volume} {11}},\ \bibinfo {pages} {e1004596}
  (\bibinfo {year} {2015})}\BibitemShut {NoStop}%
\bibitem [{\citenamefont {Savelli}\ \emph {et~al.}(2017)\citenamefont
  {Savelli}, \citenamefont {Luck},\ and\ \citenamefont
  {Knierim}}]{savelli2017framing}%
  \BibitemOpen
  \bibfield  {author} {\bibinfo {author} {\bibfnamefont {F.}~\bibnamefont
  {Savelli}}, \bibinfo {author} {\bibfnamefont {J.~D.}\ \bibnamefont {Luck}}, \
  and\ \bibinfo {author} {\bibfnamefont {J.~J.}\ \bibnamefont {Knierim}},\
  }\href@noop {} {\bibfield  {journal} {\bibinfo  {journal} {eLife}\ }\textbf
  {\bibinfo {volume} {6}} (\bibinfo {year} {2017})}\BibitemShut {NoStop}%
\bibitem [{\citenamefont {Krupic}\ \emph {et~al.}(2016)\citenamefont {Krupic},
  \citenamefont {Bauza}, \citenamefont {Burton},\ and\ \citenamefont
  {O'Keefe}}]{krupic2016framing}%
  \BibitemOpen
  \bibfield  {author} {\bibinfo {author} {\bibfnamefont {J.}~\bibnamefont
  {Krupic}}, \bibinfo {author} {\bibfnamefont {M.}~\bibnamefont {Bauza}},
  \bibinfo {author} {\bibfnamefont {S.}~\bibnamefont {Burton}}, \ and\ \bibinfo
  {author} {\bibfnamefont {J.}~\bibnamefont {O'Keefe}},\ }\href@noop {}
  {\bibfield  {journal} {\bibinfo  {journal} {J. Physiol.}\ }\textbf {\bibinfo
  {volume} {594}},\ \bibinfo {pages} {6489} (\bibinfo {year}
  {2016})}\BibitemShut {NoStop}%
\bibitem [{\citenamefont {Giocomo}(2016)}]{giocomo2016environmental}%
  \BibitemOpen
  \bibfield  {author} {\bibinfo {author} {\bibfnamefont {L.~M.}\ \bibnamefont
  {Giocomo}},\ }\href@noop {} {\bibfield  {journal} {\bibinfo  {journal} {J.
  Physiol.}\ }\textbf {\bibinfo {volume} {594}},\ \bibinfo {pages} {6501}
  (\bibinfo {year} {2016})}\BibitemShut {NoStop}%
\bibitem [{\citenamefont {Evans}\ \emph {et~al.}(2016)\citenamefont {Evans},
  \citenamefont {Bicanski}, \citenamefont {Bush},\ and\ \citenamefont
  {Burgess}}]{Evans:2016cf}%
  \BibitemOpen
  \bibfield  {author} {\bibinfo {author} {\bibfnamefont {T.}~\bibnamefont
  {Evans}}, \bibinfo {author} {\bibfnamefont {A.}~\bibnamefont {Bicanski}},
  \bibinfo {author} {\bibfnamefont {D.}~\bibnamefont {Bush}}, \ and\ \bibinfo
  {author} {\bibfnamefont {N.}~\bibnamefont {Burgess}},\ }\href@noop {}
  {\bibfield  {journal} {\bibinfo  {journal} {J. Physiol.}\ }\textbf {\bibinfo
  {volume} {594}},\ \bibinfo {pages} {6535} (\bibinfo {year}
  {2016})}\BibitemShut {NoStop}%
\bibitem [{\citenamefont {Hardcastle}\ \emph {et~al.}(2017)\citenamefont
  {Hardcastle}, \citenamefont {Maheswaranathan}, \citenamefont {Ganguli},\ and\
  \citenamefont {Giocomo}}]{Hardcastle:2017ce}%
  \BibitemOpen
  \bibfield  {author} {\bibinfo {author} {\bibfnamefont {K.}~\bibnamefont
  {Hardcastle}}, \bibinfo {author} {\bibfnamefont {N.}~\bibnamefont
  {Maheswaranathan}}, \bibinfo {author} {\bibfnamefont {S.}~\bibnamefont
  {Ganguli}}, \ and\ \bibinfo {author} {\bibfnamefont {L.~M.}\ \bibnamefont
  {Giocomo}},\ }\href@noop {} {\bibfield  {journal} {\bibinfo  {journal}
  {Neuron}\ }\textbf {\bibinfo {volume} {94}},\ \bibinfo {pages} {375}
  (\bibinfo {year} {2017})}\BibitemShut {NoStop}%
\bibitem [{\citenamefont {Keinath}\ \emph {et~al.}(2018)\citenamefont
  {Keinath}, \citenamefont {Epstein},\ and\ \citenamefont
  {Balasubramanian}}]{Keinath:2018el}%
  \BibitemOpen
  \bibfield  {author} {\bibinfo {author} {\bibfnamefont {A.~T.}\ \bibnamefont
  {Keinath}}, \bibinfo {author} {\bibfnamefont {R.~A.}\ \bibnamefont
  {Epstein}}, \ and\ \bibinfo {author} {\bibfnamefont {V.}~\bibnamefont
  {Balasubramanian}},\ }\href@noop {} {\bibfield  {journal} {\bibinfo
  {journal} {eLife}\ }\textbf {\bibinfo {volume} {7}},\ \bibinfo {pages} {71}
  (\bibinfo {year} {2018})}\BibitemShut {NoStop}%
\bibitem [{\citenamefont {Ocko}\ \emph {et~al.}(2018)\citenamefont {Ocko},
  \citenamefont {Hardcastle}, \citenamefont {Giocomo},\ and\ \citenamefont
  {Ganguli}}]{Ocko:2018ed}%
  \BibitemOpen
  \bibfield  {author} {\bibinfo {author} {\bibfnamefont {S.~A.}\ \bibnamefont
  {Ocko}}, \bibinfo {author} {\bibfnamefont {K.}~\bibnamefont {Hardcastle}},
  \bibinfo {author} {\bibfnamefont {L.~M.}\ \bibnamefont {Giocomo}}, \ and\
  \bibinfo {author} {\bibfnamefont {S.}~\bibnamefont {Ganguli}},\ }\href@noop
  {} {\bibfield  {journal} {\bibinfo  {journal} {Proc. Natl. Acad. Sci. U. S.
  A.}\ }\textbf {\bibinfo {volume} {115}},\ \bibinfo {pages} {E11798} (\bibinfo
  {year} {2018})}\BibitemShut {NoStop}%
\bibitem [{\citenamefont {Fuchs}\ \emph {et~al.}(2016)\citenamefont {Fuchs},
  \citenamefont {Neitz}, \citenamefont {Pinna}, \citenamefont {Melzer},
  \citenamefont {Caputi},\ and\ \citenamefont {Monyer}}]{Fuchs:2016et}%
  \BibitemOpen
  \bibfield  {author} {\bibinfo {author} {\bibfnamefont {E.~C.}\ \bibnamefont
  {Fuchs}}, \bibinfo {author} {\bibfnamefont {A.}~\bibnamefont {Neitz}},
  \bibinfo {author} {\bibfnamefont {R.}~\bibnamefont {Pinna}}, \bibinfo
  {author} {\bibfnamefont {S.}~\bibnamefont {Melzer}}, \bibinfo {author}
  {\bibfnamefont {A.}~\bibnamefont {Caputi}}, \ and\ \bibinfo {author}
  {\bibfnamefont {H.}~\bibnamefont {Monyer}},\ }\href@noop {} {\bibfield
  {journal} {\bibinfo  {journal} {Neuron}\ }\textbf {\bibinfo {volume} {89}},\
  \bibinfo {pages} {194} (\bibinfo {year} {2016})}\BibitemShut {NoStop}%
\bibitem [{\citenamefont {Zutshi}\ \emph {et~al.}(2018)\citenamefont {Zutshi},
  \citenamefont {Fu}, \citenamefont {Lilascharoen}, \citenamefont {Leutgeb},
  \citenamefont {Lim},\ and\ \citenamefont {Leutgeb}}]{Zutshi:2018ku}%
  \BibitemOpen
  \bibfield  {author} {\bibinfo {author} {\bibfnamefont {I.}~\bibnamefont
  {Zutshi}}, \bibinfo {author} {\bibfnamefont {M.~L.}\ \bibnamefont {Fu}},
  \bibinfo {author} {\bibfnamefont {V.}~\bibnamefont {Lilascharoen}}, \bibinfo
  {author} {\bibfnamefont {J.~K.}\ \bibnamefont {Leutgeb}}, \bibinfo {author}
  {\bibfnamefont {B.~K.}\ \bibnamefont {Lim}}, \ and\ \bibinfo {author}
  {\bibfnamefont {S.}~\bibnamefont {Leutgeb}},\ }\href@noop {} {\bibfield
  {journal} {\bibinfo  {journal} {Nat. Commun.}\ }\textbf {\bibinfo {volume}
  {9}},\ \bibinfo {pages} {3701} (\bibinfo {year} {2018})}\BibitemShut
  {NoStop}%
\bibitem [{\citenamefont {Mathis}\ \emph {et~al.}(2013)\citenamefont {Mathis},
  \citenamefont {Herz},\ and\ \citenamefont {Stemmler}}]{mathis2013multiscale}%
  \BibitemOpen
  \bibfield  {author} {\bibinfo {author} {\bibfnamefont {A.}~\bibnamefont
  {Mathis}}, \bibinfo {author} {\bibfnamefont {A.~V.~M.}\ \bibnamefont {Herz}},
  \ and\ \bibinfo {author} {\bibfnamefont {M.~B.}\ \bibnamefont {Stemmler}},\
  }\href@noop {} {\bibfield  {journal} {\bibinfo  {journal} {Physical Review
  E}\ }\textbf {\bibinfo {volume} {88}},\ \bibinfo {pages} {022713} (\bibinfo
  {year} {2013})}\BibitemShut {NoStop}%
\bibitem [{\citenamefont {Tocker}\ \emph {et~al.}(2015)\citenamefont {Tocker},
  \citenamefont {Barak},\ and\ \citenamefont {Derdikman}}]{Tocker:2015ff}%
  \BibitemOpen
  \bibfield  {author} {\bibinfo {author} {\bibfnamefont {G.}~\bibnamefont
  {Tocker}}, \bibinfo {author} {\bibfnamefont {O.}~\bibnamefont {Barak}}, \
  and\ \bibinfo {author} {\bibfnamefont {D.}~\bibnamefont {Derdikman}},\
  }\href@noop {} {\bibfield  {journal} {\bibinfo  {journal} {Hippocampus}\
  }\textbf {\bibinfo {volume} {25}},\ \bibinfo {pages} {1599} (\bibinfo {year}
  {2015})}\BibitemShut {NoStop}%
\bibitem [{\citenamefont {Heys}\ \emph {et~al.}(2014)\citenamefont {Heys},
  \citenamefont {Rangarajan},\ and\ \citenamefont {Dombeck}}]{Heys:2014cv}%
  \BibitemOpen
  \bibfield  {author} {\bibinfo {author} {\bibfnamefont {J.~G.}\ \bibnamefont
  {Heys}}, \bibinfo {author} {\bibfnamefont {K.~V.}\ \bibnamefont
  {Rangarajan}}, \ and\ \bibinfo {author} {\bibfnamefont {D.~A.}\ \bibnamefont
  {Dombeck}},\ }\href@noop {} {\bibfield  {journal} {\bibinfo  {journal}
  {Neuron}\ }\textbf {\bibinfo {volume} {84}},\ \bibinfo {pages} {1079}
  (\bibinfo {year} {2014})}\BibitemShut {NoStop}%
\bibitem [{\citenamefont {Gu}\ \emph {et~al.}(2018)\citenamefont {Gu},
  \citenamefont {Lewallen}, \citenamefont {Kinkhabwala}, \citenamefont
  {Domnisoru}, \citenamefont {Yoon}, \citenamefont {Gauthier}, \citenamefont
  {Fiete},\ and\ \citenamefont {Tank}}]{Gu:2018dm}%
  \BibitemOpen
  \bibfield  {author} {\bibinfo {author} {\bibfnamefont {Y.}~\bibnamefont
  {Gu}}, \bibinfo {author} {\bibfnamefont {S.}~\bibnamefont {Lewallen}},
  \bibinfo {author} {\bibfnamefont {A.~A.}\ \bibnamefont {Kinkhabwala}},
  \bibinfo {author} {\bibfnamefont {C.}~\bibnamefont {Domnisoru}}, \bibinfo
  {author} {\bibfnamefont {K.}~\bibnamefont {Yoon}}, \bibinfo {author}
  {\bibfnamefont {J.~L.}\ \bibnamefont {Gauthier}}, \bibinfo {author}
  {\bibfnamefont {I.~R.}\ \bibnamefont {Fiete}}, \ and\ \bibinfo {author}
  {\bibfnamefont {D.~W.}\ \bibnamefont {Tank}},\ }\href@noop {} {\bibfield
  {journal} {\bibinfo  {journal} {Cell}\ }\textbf {\bibinfo {volume} {175}},\
  \bibinfo {pages} {736} (\bibinfo {year} {2018})}\BibitemShut {NoStop}%
\bibitem [{\citenamefont {Couey}\ \emph {et~al.}(2013)\citenamefont {Couey},
  \citenamefont {Witoelar}, \citenamefont {Zhang}, \citenamefont {Zheng},
  \citenamefont {Ye}, \citenamefont {Dunn}, \citenamefont {Czajkowski},
  \citenamefont {Moser}, \citenamefont {Moser}, \citenamefont {Roudi},\ and\
  \citenamefont {Witter}}]{Couey:2013fi}%
  \BibitemOpen
  \bibfield  {author} {\bibinfo {author} {\bibfnamefont {J.~J.}\ \bibnamefont
  {Couey}}, \bibinfo {author} {\bibfnamefont {A.}~\bibnamefont {Witoelar}},
  \bibinfo {author} {\bibfnamefont {S.-J.}\ \bibnamefont {Zhang}}, \bibinfo
  {author} {\bibfnamefont {K.}~\bibnamefont {Zheng}}, \bibinfo {author}
  {\bibfnamefont {J.}~\bibnamefont {Ye}}, \bibinfo {author} {\bibfnamefont
  {B.}~\bibnamefont {Dunn}}, \bibinfo {author} {\bibfnamefont {R.}~\bibnamefont
  {Czajkowski}}, \bibinfo {author} {\bibfnamefont {M.-B.}\ \bibnamefont
  {Moser}}, \bibinfo {author} {\bibfnamefont {E.~I.}\ \bibnamefont {Moser}},
  \bibinfo {author} {\bibfnamefont {Y.}~\bibnamefont {Roudi}}, \ and\ \bibinfo
  {author} {\bibfnamefont {M.~P.}\ \bibnamefont {Witter}},\ }\href@noop {}
  {\bibfield  {journal} {\bibinfo  {journal} {Nat. Neurosci.}\ }\textbf
  {\bibinfo {volume} {16}},\ \bibinfo {pages} {318} (\bibinfo {year}
  {2013})}\BibitemShut {NoStop}%
\bibitem [{\citenamefont {Winterer}\ \emph {et~al.}(2017)\citenamefont
  {Winterer}, \citenamefont {Maier}, \citenamefont {Wozny}, \citenamefont
  {Beed}, \citenamefont {Breustedt}, \citenamefont {Evangelista}, \citenamefont
  {Peng}, \citenamefont {D{\textquoteright}Albis}, \citenamefont {Kempter},\
  and\ \citenamefont {Schmitz}}]{Winterer:2017fl}%
  \BibitemOpen
  \bibfield  {author} {\bibinfo {author} {\bibfnamefont {J.}~\bibnamefont
  {Winterer}}, \bibinfo {author} {\bibfnamefont {N.}~\bibnamefont {Maier}},
  \bibinfo {author} {\bibfnamefont {C.}~\bibnamefont {Wozny}}, \bibinfo
  {author} {\bibfnamefont {P.}~\bibnamefont {Beed}}, \bibinfo {author}
  {\bibfnamefont {J.}~\bibnamefont {Breustedt}}, \bibinfo {author}
  {\bibfnamefont {R.}~\bibnamefont {Evangelista}}, \bibinfo {author}
  {\bibfnamefont {Y.}~\bibnamefont {Peng}}, \bibinfo {author} {\bibfnamefont
  {T.}~\bibnamefont {D{\textquoteright}Albis}}, \bibinfo {author}
  {\bibfnamefont {R.}~\bibnamefont {Kempter}}, \ and\ \bibinfo {author}
  {\bibfnamefont {D.}~\bibnamefont {Schmitz}},\ }\href@noop {} {\bibfield
  {journal} {\bibinfo  {journal} {Cell Reports}\ }\textbf {\bibinfo {volume}
  {19}},\ \bibinfo {pages} {1110} (\bibinfo {year} {2017})}\BibitemShut
  {NoStop}%
\bibitem [{\citenamefont {Widloski}\ and\ \citenamefont
  {Fiete}(2017)}]{widloskifiete}%
  \BibitemOpen
  \bibfield  {author} {\bibinfo {author} {\bibfnamefont {J.}~\bibnamefont
  {Widloski}}\ and\ \bibinfo {author} {\bibfnamefont {I.~R.}\ \bibnamefont
  {Fiete}},\ }\href@noop {} {}\bibinfo {howpublished} {Personal communication}
  (\bibinfo {year} {2017})\BibitemShut {NoStop}%
\bibitem [{\citenamefont {Krupic}\ \emph {et~al.}(2012)\citenamefont {Krupic},
  \citenamefont {Burgess},\ and\ \citenamefont {O'Keefe}}]{Krupic:2012id}%
  \BibitemOpen
  \bibfield  {author} {\bibinfo {author} {\bibfnamefont {J.}~\bibnamefont
  {Krupic}}, \bibinfo {author} {\bibfnamefont {N.}~\bibnamefont {Burgess}}, \
  and\ \bibinfo {author} {\bibfnamefont {J.}~\bibnamefont {O'Keefe}},\
  }\href@noop {} {\bibfield  {journal} {\bibinfo  {journal} {Science}\ }\textbf
  {\bibinfo {volume} {337}},\ \bibinfo {pages} {853} (\bibinfo {year}
  {2012})}\BibitemShut {NoStop}%
\bibitem [{\citenamefont {Navratilova}\ \emph {et~al.}(2016)\citenamefont
  {Navratilova}, \citenamefont {Godfrey},\ and\ \citenamefont
  {McNaughton}}]{Navratilova:2016iq}%
  \BibitemOpen
  \bibfield  {author} {\bibinfo {author} {\bibfnamefont {Z.}~\bibnamefont
  {Navratilova}}, \bibinfo {author} {\bibfnamefont {K.~B.}\ \bibnamefont
  {Godfrey}}, \ and\ \bibinfo {author} {\bibfnamefont {B.~L.}\ \bibnamefont
  {McNaughton}},\ }\href@noop {} {\bibfield  {journal} {\bibinfo  {journal} {J.
  Neurophysiol.}\ }\textbf {\bibinfo {volume} {115}},\ \bibinfo {pages} {992}
  (\bibinfo {year} {2016})}\BibitemShut {NoStop}%
\bibitem [{\citenamefont {Dunn}\ \emph {et~al.}(2015)\citenamefont {Dunn},
  \citenamefont {M{\o}rreaunet},\ and\ \citenamefont {Roudi}}]{Dunn:2015he}%
  \BibitemOpen
  \bibfield  {author} {\bibinfo {author} {\bibfnamefont {B.}~\bibnamefont
  {Dunn}}, \bibinfo {author} {\bibfnamefont {M.}~\bibnamefont {M{\o}rreaunet}},
  \ and\ \bibinfo {author} {\bibfnamefont {Y.}~\bibnamefont {Roudi}},\
  }\href@noop {} {\bibfield  {journal} {\bibinfo  {journal} {PLOS Comp. Biol.}\
  }\textbf {\bibinfo {volume} {11}},\ \bibinfo {pages} {e1004052} (\bibinfo
  {year} {2015})}\BibitemShut {NoStop}%
\bibitem [{\citenamefont {Yoon}\ \emph {et~al.}(2013)\citenamefont {Yoon},
  \citenamefont {Buice}, \citenamefont {Barry}, \citenamefont {Hayman},
  \citenamefont {Burgess},\ and\ \citenamefont {Fiete}}]{Yoon:2013hv}%
  \BibitemOpen
  \bibfield  {author} {\bibinfo {author} {\bibfnamefont {K.}~\bibnamefont
  {Yoon}}, \bibinfo {author} {\bibfnamefont {M.~A.}\ \bibnamefont {Buice}},
  \bibinfo {author} {\bibfnamefont {C.}~\bibnamefont {Barry}}, \bibinfo
  {author} {\bibfnamefont {R.}~\bibnamefont {Hayman}}, \bibinfo {author}
  {\bibfnamefont {N.}~\bibnamefont {Burgess}}, \ and\ \bibinfo {author}
  {\bibfnamefont {I.~R.}\ \bibnamefont {Fiete}},\ }\href@noop {} {\bibfield
  {journal} {\bibinfo  {journal} {Nat. Neurosci.}\ }\textbf {\bibinfo {volume}
  {16}},\ \bibinfo {pages} {1077} (\bibinfo {year} {2013})}\BibitemShut
  {NoStop}%
\bibitem [{\citenamefont {Kontorova}\ and\ \citenamefont
  {Frenkel}(1938)}]{Kontorova:1938en}%
  \BibitemOpen
  \bibfield  {author} {\bibinfo {author} {\bibfnamefont {T.}~\bibnamefont
  {Kontorova}}\ and\ \bibinfo {author} {\bibfnamefont {J.}~\bibnamefont
  {Frenkel}},\ }\href@noop {} {\bibfield  {journal} {\bibinfo  {journal} {Zh.
  Eksp. Teor. Fiz.}\ }\textbf {\bibinfo {volume} {8}},\ \bibinfo {pages} {1340}
  (\bibinfo {year} {1938})}\BibitemShut {NoStop}%
\bibitem [{\citenamefont {Bak}(1982)}]{Bak:1982it}%
  \BibitemOpen
  \bibfield  {author} {\bibinfo {author} {\bibfnamefont {P.}~\bibnamefont
  {Bak}},\ }\href@noop {} {\bibfield  {journal} {\bibinfo  {journal} {Rep.
  Prog. Phys.}\ }\textbf {\bibinfo {volume} {45}},\ \bibinfo {pages} {587}
  (\bibinfo {year} {1982})}\BibitemShut {NoStop}%
\bibitem [{\citenamefont {Levine}\ and\ \citenamefont
  {Steinhardt}(1986)}]{Levine:1986kb}%
  \BibitemOpen
  \bibfield  {author} {\bibinfo {author} {\bibfnamefont {D.}~\bibnamefont
  {Levine}}\ and\ \bibinfo {author} {\bibfnamefont {P.~J.}\ \bibnamefont
  {Steinhardt}},\ }\href@noop {} {\bibfield  {journal} {\bibinfo  {journal}
  {Phys. Rev. B}\ }\textbf {\bibinfo {volume} {34}},\ \bibinfo {pages} {596}
  (\bibinfo {year} {1986})}\BibitemShut {NoStop}%
\bibitem [{\citenamefont {F{\"o}rster}\ \emph {et~al.}(2013)\citenamefont
  {F{\"o}rster}, \citenamefont {Meinel}, \citenamefont {Hammer}, \citenamefont
  {Trautmann},\ and\ \citenamefont {Widdra}}]{Forster:2013de}%
  \BibitemOpen
  \bibfield  {author} {\bibinfo {author} {\bibfnamefont {S.}~\bibnamefont
  {F{\"o}rster}}, \bibinfo {author} {\bibfnamefont {K.}~\bibnamefont {Meinel}},
  \bibinfo {author} {\bibfnamefont {R.}~\bibnamefont {Hammer}}, \bibinfo
  {author} {\bibfnamefont {M.}~\bibnamefont {Trautmann}}, \ and\ \bibinfo
  {author} {\bibfnamefont {W.}~\bibnamefont {Widdra}},\ }\href@noop {}
  {\bibfield  {journal} {\bibinfo  {journal} {Nature}\ }\textbf {\bibinfo
  {volume} {502}},\ \bibinfo {pages} {215} (\bibinfo {year}
  {2013})}\BibitemShut {NoStop}%
\bibitem [{\citenamefont {F{\"o}rster}\ \emph {et~al.}(2016)\citenamefont
  {F{\"o}rster}, \citenamefont {Trautmann}, \citenamefont {Roy}, \citenamefont
  {Adeagbo}, \citenamefont {Zollner}, \citenamefont {Hammer}, \citenamefont
  {Schumann}, \citenamefont {Meinel}, \citenamefont {Nayak}, \citenamefont
  {Mohseni}, \citenamefont {Hergert}, \citenamefont {Meyerheim},\ and\
  \citenamefont {Widdra}}]{Forster:2016bd}%
  \BibitemOpen
  \bibfield  {author} {\bibinfo {author} {\bibfnamefont {S.}~\bibnamefont
  {F{\"o}rster}}, \bibinfo {author} {\bibfnamefont {M.}~\bibnamefont
  {Trautmann}}, \bibinfo {author} {\bibfnamefont {S.}~\bibnamefont {Roy}},
  \bibinfo {author} {\bibfnamefont {W.~A.}\ \bibnamefont {Adeagbo}}, \bibinfo
  {author} {\bibfnamefont {E.~M.}\ \bibnamefont {Zollner}}, \bibinfo {author}
  {\bibfnamefont {R.}~\bibnamefont {Hammer}}, \bibinfo {author} {\bibfnamefont
  {F.~O.}\ \bibnamefont {Schumann}}, \bibinfo {author} {\bibfnamefont
  {K.}~\bibnamefont {Meinel}}, \bibinfo {author} {\bibfnamefont {S.~K.}\
  \bibnamefont {Nayak}}, \bibinfo {author} {\bibfnamefont {K.}~\bibnamefont
  {Mohseni}}, \bibinfo {author} {\bibfnamefont {W.}~\bibnamefont {Hergert}},
  \bibinfo {author} {\bibfnamefont {H.~L.}\ \bibnamefont {Meyerheim}}, \ and\
  \bibinfo {author} {\bibfnamefont {W.}~\bibnamefont {Widdra}},\ }\href@noop {}
  {\bibfield  {journal} {\bibinfo  {journal} {Phys. Rev. Lett.}\ }\textbf
  {\bibinfo {volume} {117}},\ \bibinfo {pages} {1260} (\bibinfo {year}
  {2016})}\BibitemShut {NoStop}%
\bibitem [{\citenamefont {Pa{\ss}ens}\ \emph {et~al.}(2017)\citenamefont
  {Pa{\ss}ens}, \citenamefont {Caciuc}, \citenamefont {Atodiresei},
  \citenamefont {Feuerbacher}, \citenamefont {Moors}, \citenamefont
  {Dunin-Borkowski}, \citenamefont {Bl{\"u}gel}, \citenamefont {Waser},\ and\
  \citenamefont {Karth{\"a}user}}]{Passens:2017kl}%
  \BibitemOpen
  \bibfield  {author} {\bibinfo {author} {\bibfnamefont {M.}~\bibnamefont
  {Pa{\ss}ens}}, \bibinfo {author} {\bibfnamefont {V.}~\bibnamefont {Caciuc}},
  \bibinfo {author} {\bibfnamefont {N.}~\bibnamefont {Atodiresei}}, \bibinfo
  {author} {\bibfnamefont {M.}~\bibnamefont {Feuerbacher}}, \bibinfo {author}
  {\bibfnamefont {M.}~\bibnamefont {Moors}}, \bibinfo {author} {\bibfnamefont
  {R.~E.}\ \bibnamefont {Dunin-Borkowski}}, \bibinfo {author} {\bibfnamefont
  {S.}~\bibnamefont {Bl{\"u}gel}}, \bibinfo {author} {\bibfnamefont
  {R.}~\bibnamefont {Waser}}, \ and\ \bibinfo {author} {\bibfnamefont
  {S.}~\bibnamefont {Karth{\"a}user}},\ }\href@noop {} {\bibfield  {journal}
  {\bibinfo  {journal} {Nat. Commun.}\ }\textbf {\bibinfo {volume} {8}},\
  \bibinfo {pages} {15367} (\bibinfo {year} {2017})}\BibitemShut {NoStop}%
\bibitem [{\citenamefont {Widloski}\ and\ \citenamefont
  {Fiete}(2014)}]{Widloski:2014dl}%
  \BibitemOpen
  \bibfield  {author} {\bibinfo {author} {\bibfnamefont {J.}~\bibnamefont
  {Widloski}}\ and\ \bibinfo {author} {\bibfnamefont {I.~R.}\ \bibnamefont
  {Fiete}},\ }\href@noop {} {\bibfield  {journal} {\bibinfo  {journal}
  {Neuron}\ }\textbf {\bibinfo {volume} {83}},\ \bibinfo {pages} {481}
  (\bibinfo {year} {2014})}\BibitemShut {NoStop}%
\bibitem [{\citenamefont {Hardcastle}\ \emph {et~al.}(2015)\citenamefont
  {Hardcastle}, \citenamefont {Ganguli},\ and\ \citenamefont
  {Giocomo}}]{Hardcastle:2015il}%
  \BibitemOpen
  \bibfield  {author} {\bibinfo {author} {\bibfnamefont {K.}~\bibnamefont
  {Hardcastle}}, \bibinfo {author} {\bibfnamefont {S.}~\bibnamefont {Ganguli}},
  \ and\ \bibinfo {author} {\bibfnamefont {L.~M.}\ \bibnamefont {Giocomo}},\
  }\href@noop {} {\bibfield  {journal} {\bibinfo  {journal} {Neuron}\ }\textbf
  {\bibinfo {volume} {86}},\ \bibinfo {pages} {827} (\bibinfo {year}
  {2015})}\BibitemShut {NoStop}%
\bibitem [{\citenamefont {Pollock}\ \emph {et~al.}(2017)\citenamefont
  {Pollock}, \citenamefont {Desai}, \citenamefont {Wei},\ and\ \citenamefont
  {Balasubramanian}}]{pollock}%
  \BibitemOpen
  \bibfield  {author} {\bibinfo {author} {\bibfnamefont {E.}~\bibnamefont
  {Pollock}}, \bibinfo {author} {\bibfnamefont {N.}~\bibnamefont {Desai}},
  \bibinfo {author} {\bibfnamefont {X.}~\bibnamefont {Wei}}, \ and\ \bibinfo
  {author} {\bibfnamefont {V.}~\bibnamefont {Balasubramanian}},\ }in\
  \href@noop {} {\emph {\bibinfo {booktitle} {Cosyne Abstracts 2017}}}\
  (\bibinfo {address} {Salt Lake City, UT, USA},\ \bibinfo {year}
  {2017})\BibitemShut {NoStop}%
\bibitem [{\citenamefont {Mosheiff}\ and\ \citenamefont
  {Burak}(2018)}]{mosheiffburak}%
  \BibitemOpen
  \bibfield  {author} {\bibinfo {author} {\bibfnamefont {N.}~\bibnamefont
  {Mosheiff}}\ and\ \bibinfo {author} {\bibfnamefont {Y.}~\bibnamefont
  {Burak}},\ }in\ \href@noop {} {\emph {\bibinfo {booktitle} {Cosyne Abstracts
  2018}}}\ (\bibinfo {address} {Denver, CO, USA},\ \bibinfo {year}
  {2018})\BibitemShut {NoStop}%
\bibitem [{\citenamefont {Weber}\ and\ \citenamefont
  {Sprekeler}(2019)}]{Weber:2019gb}%
  \BibitemOpen
  \bibfield  {author} {\bibinfo {author} {\bibfnamefont {S.~N.}\ \bibnamefont
  {Weber}}\ and\ \bibinfo {author} {\bibfnamefont {H.}~\bibnamefont
  {Sprekeler}},\ }\href@noop {} {\bibfield  {journal} {\bibinfo  {journal}
  {PLOS Comp. Biol.}\ }\textbf {\bibinfo {volume} {15}},\ \bibinfo {pages}
  {e1006804} (\bibinfo {year} {2019})}\BibitemShut {NoStop}%
\end{thebibliography}
\end{document}